%% file: qg-paper.tex
\documentclass[twocolumn,epjc3]{svjour3}

\RequirePackage{flushend}
\RequirePackage[numbers,sort&compress]{natbib}
\RequirePackage[colorlinks,citecolor=blue,urlcolor=blue,linkcolor=blue,hyperindex,breaklinks]{hyperref}

\usepackage{graphicx}
\usepackage{dcolumn}
\usepackage{bm}

\usepackage{subfigure}
\usepackage{amsmath}
\usepackage{amssymb}
\usepackage{multirow}
\usepackage{rotating}
\usepackage{slashed}
\usepackage{url}
\usepackage{xspace}
\usepackage{comment}
\usepackage{multirow}
\usepackage{etoolbox}

\newcommand{\ie}{\text{i.e.}\xspace}

\def\TeV{\ifmmode {\mathrm{\ Te\kern -0.1em V}}\else
                   \textrm{Te\kern -0.1em V}\fi}%
\def\GeV{\ifmmode {\mathrm{\ Ge\kern -0.1em V}}\else
                   \textrm{Ge\kern -0.1em V}\fi}%
\def\MeV{\ifmmode {\mathrm{\ Me\kern -0.1em V}}\else
                   \textrm{Me\kern -0.1em V}\fi}%
\def\pt{\ensuremath{p_{\mathrm{T}}}} 
\def\ET{\ensuremath{E_{\mathrm{T}}}} 
\def\mt{\ensuremath{m_{\mathrm{T}}}} 
\def\Wboson{\ensuremath{W}}%
\def\Zboson{\ensuremath{Z}}
\def\antibar#1{\ensuremath{#1\bar{#1}}}

\def\ttbar{\antibar{t}}

\def\qqbar{\antibar{q}}
\def\MET{\ensuremath{E_{\mathrm{T}}^{\mathrm{miss}}}} 

\def\bbbar{\antibar{b}}

\def\ccbar{\antibar{c}}
\def\sbar{\ensuremath{\bar{s}}}
\newcommand{\AKT}{anti-\ensuremath{k_{t}}\xspace}
\newcommand{\ptjet}{\ensuremath{\pt^{\rm \ jet}}\xspace}

\newcommand{\etajet}{\ensuremath{\eta^{\rm jet}}\xspace}
\newcommand{\phijet}{\ensuremath{\phi^{\rm \ jet}}\xspace}

\newcommand{\DeltaR}{\ensuremath{\Delta R}\xspace}

\newcommand{\gjet}{\ensuremath{\gamma\text{+jet}}\xspace}

\usepackage{preprintcover}  

\PreprintCoverPaperTitle{Light-quark and gluon jet discrimination in $pp$ collisions at $\sqrt{s}=7\TeV$ with the ATLAS detector}  

\PreprintIdNumber{CERN-PH-EP-2014-058}  

\PreprintCoverAbstract{
A likelihood-based discriminant for the identification of quark- and gluon-initiated jets is built and validated using
4.7~fb$^{-1}$ of proton--proton collision data at $\sqrt{s}=7$~\TeV\ collected with the ATLAS detector at the LHC.
Data samples with enriched quark or gluon content are used in the construction and validation of
templates of jet properties that are the input to the likelihood-based discriminant.
The discriminating power of the jet tagger is
established in both data and Monte Carlo samples within a systematic uncertainty of
$\approx 10$--$20\%$. In data, light-quark jets can be tagged with an efficiency of $\approx 50\%$
while achieving a gluon-jet mis-tag rate
of $\approx 25\%$ in a $\pt$ range between $40\GeV$ and
$360\GeV$ for jets in the acceptance of the tracker. The rejection of gluon-jets found
in the data is significantly below what is
attainable using a {\sc Pythia}~6 Monte Carlo simulation, where gluon-jet mis-tag rates of 10\% can be
reached for a 50\% selection efficiency of light-quark jets using the same jet properties.
}  

\PreprintJournalName{Eur. Phys. J. C}  

\begin{document}


\title{Light-quark and gluon jet discrimination in $pp$ collisions at $\sqrt{s}=7\TeV$ with the ATLAS detector}

\titlerunning{Light-quark and gluon jet discrimiation with ATLAS}        

\author{The ATLAS Collaboration}

\institute{}

\date{Received: date / Accepted: date}

\maketitle

\begin{abstract}
A likelihood-based discriminant for the identification of quark- and gluon-initiated jets is built and validated using 
4.7~fb$^{-1}$ of proton--proton collision data at $\sqrt{s}=7$~\TeV\ collected with the ATLAS detector at the LHC.
Data samples with enriched quark or gluon content are used in the construction and validation of  
templates of jet properties that are the input to the likelihood-based discriminant. 
The discriminating power of the jet tagger is
established in both data and Monte Carlo samples within a systematic uncertainty of 
$\approx 10$--$20\%$. In data, light-quark jets can be tagged with an efficiency of $\approx 50\%$ 
while achieving a gluon-jet mis-tag rate 
of $\approx 25\%$ in a $\pt$ range between $40\GeV$ and
$360\GeV$ for jets in the acceptance of the tracker. The rejection of gluon-jets found 
in the data is significantly below what is 
attainable using a {\sc Pythia}~6 Monte Carlo simulation, where gluon-jet mis-tag rates of 10\% can be  
reached for a 50\% selection efficiency of light-quark jets using the same jet properties. 
\end{abstract}

\section{Introduction}
\label{sec:intro}


The production of quarks and gluons via strong interactions is the dominant high-momentum-transfer
process
at the LHC and is a significant background to most new-physics searches.
These partons are measured as jets, which are collimated streams of charged and neutral particles, clustered using dedicated algorithms.
Corrections to measured quantities are necessary to relate the jets to their
parent partons. Many gluons are generated in most common Standard Model processes, such as the inclusive production of
jets~\cite{InclJets,Aad:2011tqa}. On the other hand,
some processes arising from new-physics models, for example supersymmetry, generate many light quarks~\cite{ATLASSUSY1,ATLASSUSY2}.
The power to discriminate between jets initiated by light quarks and those initiated
by gluons would therefore provide a powerful tool in searches for new physics.
In case of a discovery of a new particle, such a discriminant could provide
valuable information about its nature.
Also, some Standard Model measurements rely on the correct identification of the origin of jets,
as in the cases of reconstructing a hadronic $\Wboson$ decay when measuring the top quark mass, or in the reconstruction
of a hadronic $\Zboson$ decay when measuring the Higgs boson mass via $h\rightarrow ZZ\rightarrow\ell\ell \qqbar$.
These analyses would benefit from such a discriminant. These applications motivate the analysis of the partonic
origin of jets that is the focus of this paper.

In perturbative quantum chromodynamics (QCD), the concept of a parton initiating a
jet is a fixed-order notion. In the matrix-element
calculation of a high-momentum-transfer-process, the outgoing partons appear na\"ively much like outgoing
particles in the final state. However, only colourless states with two or more partons can
form an observable jet. Moreover, in a parton shower, the leading parton is only
well defined for a fixed number of splittings. The next step in the shower may change
the energy, direction, or flavour of the leading parton. Thus, labelling jets with a specific flavour
and interpreting results after such labelling requires a clearly defined procedure~\cite{schwartz1}.

Certain parton branchings can yield an ambiguous jet identity.
The labelling of a jet may also depend on the physics goal of the analysis.
For example, a jet from the $\qqbar'$ decay of a high-momentum $\Wboson$ boson produced in
a top quark decay can be considered either as a part of a top-quark jet or as a boosted
$\Wboson$-boson jet.
Nonetheless, many event topologies lend themselves to the identification of a jet as having
originated from a specific type of parton in the matrix-element calculation. Such an approach
can lead to an unambiguous and meaningful parton labelling for a large majority of jets.
This approach of linking jet-by-jet labelling to the results of the underlying leading-order (LO)
calculation is also used in this paper to define the flavour of a jet.

Discrimination between jets of different partonic origin has been attempted previously at several
experiments~\cite{TevatronShapes1,QGNN,Pumplin,QGopal,QGsub,QGlep,QGcleo,DelphiQG,DelphiQG2,AlephQG,L3QG}.
Most work has relied on jet properties that result from the difference in colour charge between
the partons. The colour factors in quantum chromodynamics differ for quarks ($C_F=4/3$) and gluons
($C_A=3$), and therefore, for example, one expects approximately $C_A/C_F=9/4$ times more particles
in a gluon-initiated jet than in a jet initiated by a light ($u$, $d$ or $s$) quark.
The measured difference in particle multiplicity at OPAL was, in fact, not far from this expectation~\cite{QGopal}.
Because of the showering that produces these additional particles, gluon jets are also expected
to be wider and have a softer particle spectrum.

The most successful studies of discrimination between light-quark-initiated and gluon-initiated jets
(henceforth, quark-jets and gluon-jets) have taken place at electron-positron colliders~\cite{QGinee1,QGinee2}.
The selection and identification of ``pure'' samples of quark- and
gluon-jets is considerably more difficult at hadron colliders because of the complication added
by beam remnants, initial-state radiation, and multi-parton interactions. The presence 
of multiple soft $pp$ collisions overlaying the hard-scatter interaction of interest at the LHC further
complicates this task. 
Recently some effort has been devoted to developing kinematic selections that significantly
enhance the fraction of quark-jets or gluon-jets in a set of events~\cite{schwartz1}.
In addition, discriminants based on jet structure have shown some promise for
distinguishing between classes of jets at the LHC~\cite{schwartz2}.

Jets that include, or are initiated by, heavy quarks (bottom and charm) also exhibit properties 
different from those of quark-jets~\cite{TevatronHQ,ATLASbjetshapes}. Generally, these jets are wider than
quark-jets. They are often identified by long-lived or leptonically decaying hadrons.
However, no special discriminant for them is developed here.

This paper is organised as follows. The ATLAS detector is briefly described in Sec.~\ref{sec:atlas}.
Section~\ref{sec:data-selection} describes details of the data and Monte Carlo (MC) samples
used, as well as the object reconstruction and event selection. Section~\ref{sec:jetdef} introduces
the definition of gluon-jets and quark-jets that are used in the remainder of the paper.
The jet properties used to build a discriminant from samples with different purities,
and the validation of the extraction method 
using MC event samples, are described in Sec.~\ref{sec:mcTemp}. 
Section~\ref{sec:disc} describes the selection of samples based on kinematic variables 
to enhance quark-jet or gluon-jet fractions and the validation of the extracted properties using
those samples.
The likelihood-based discriminant is described in Sec.~\ref{sec:perfDetails}, where its performance
in MC simulation and in data is discussed.
Finally, the conclusions are presented in Sec.~\ref{sec:conclusions}.

\section{ATLAS detector}
\label{sec:atlas}


The ATLAS detector~\cite{techpaper} comprises an inner tracking detector, a calorimeter system, and
a muon spectrometer.
The inner detector (ID) includes a silicon pixel detector, a silicon microstrip detector and a transition radiation tracker.
It is immersed in a 2~T axial magnetic field provided by a solenoid and precisely measures the trajectories of
charged particles with $|\eta|<2.5$~\footnote{ATLAS uses a right-handed coordinate system with its
origin at the nominal interaction point (IP) in the centre of the detector and the $z$-axis along the
beam pipe.
The $x$-axis points from the IP to the centre of the LHC ring, and the $y$-axis points upward.
Cylindrical coordinates $(r,\phi)$ are used in the transverse plane, $\phi$ being the azimuthal angle
around the beam pipe.
The pseudorapidity is defined in terms of the polar angle $\theta$ as $\eta=-\ln\tan(\theta/2)$, and
the rapidity is defined as $\frac{1}{2}\ln{\left(\frac{E+p_Z}{E-p_Z}\right)}$, where $E$ is the object's energy and $p_z$ is
its momentum along the $z$-axis.  The values of $\eta$, $\phi$, and $y$ are determined at the interaction vertex.
The variable $\DeltaR=\sqrt{(\Delta\eta)^2+(\Delta\phi)^2}$ is used to characterise the
angular difference between two objects using their $\eta$ and $\phi$ directions.}.
The calorimeter system covers the region $|\eta|<4.9$ and is divided into electromagnetic and hadronic
compartments.
Electromagnetic calorimetry in the region $|\eta|<3.2$ is provided by liquid-argon
sampling calorimeters with lead absorbers.
In the barrel region ($|\eta|<1.7$), the hadronic calorimeter comprises scintillator tiles with steel
absorbers, and the endcap region ($1.4<|\eta|<3.2$) is covered by a liquid-argon and copper
sampling hadronic calorimeter.
The calorimetry in the forward region ($3.2<|\eta|<4.9$) is provided by a liquid-argon and copper sampling
electromagnetic calorimeter and a liquid-argon and tungsten sampling hadronic calorimeter.
The muon spectrometer (MS) covers $|\eta|<2.7$ and uses a system of air-core toroidal magnets.

ATLAS has a three-level trigger system to select events.
The first-level trigger uses custom-built hardware components and
identifies jet, electron and photon candidates using coarse calorimeter information,
and muon candidates using coarse tracking information from the muon spectrometer.
At the highest level, full event reconstruction, similar to that used in the offline software, is performed
to accurately identify and measure objects that determine whether the event is recorded.

\section{Data sample and event selection}
\label{sec:data-selection}


Several samples are used in the construction and validation of the variables entering the
quark/gluon discriminant: dijet events, trijet events, $\gamma$+jet events,
$\gamma$+2-jet events, $t\bar{t}$ events and $W$+jet events.
After basic data quality requirements are imposed to remove known detector errors and readout problems,
the selected dataset corresponds to a total integrated luminosity of 4.67$\pm$0.08~fb$^{-1}$~\cite{NewLumiPaper}.
The data were collected from March to October 2011
at a centre-of-mass energy $\sqrt{s}=7\TeV$.
The average number of additional $pp$ collisions per bunch crossing, called
``pile-up'', rose during the data-taking period from a few to 15.

\subsection{Monte Carlo simulation}\label{sec:mc}

Simulated event samples are generated for comparison with data and
for the determination of the systematic uncertainties based on variations in
the MC generator settings. For the MC samples, several different generators are used.
{\sc MadGraph}~\cite{MadGraph} is run as a $2\rightarrow N$ generator with MLM matching~\cite{MLM}, uses
the CTEQ 6L1 parton distribution function (PDF) set, and is interfaced to {\sc Pythia} 6 with a version
of the ATLAS {\sc MC11} Underlying Event Tune 2B ({\sc AUET2B})~\cite{ATL-PHYS-PUB-2011-009}
constructed for this PDF set.  {\sc Herwig}++~\cite{Herwigpp} is run standalone
as a $2\rightarrow 2$ generator and uses the MRST LO** PDF set with the {\sc LHC-UE7-2}
tune~\cite{Gieseke:2011xy}. This tune of {\sc Herwig}++ has an improved description
of colour reconnection in multiple parton interactions and has been shown to have fair agreement
with ATLAS data in minimum-bias observables~\cite{Gieseke:2011xy}.
{\sc Pythia} 6 is also run standalone as a $2\rightarrow 2$ generator with the MRST LO**
PDF set and the {\sc AUET2B} tune.
The {\sc AUET2B} tune incorporates ATLAS~\cite{ATLASshape} and CDF~\cite{CDFshape} jet-shape measurements as
well as ATLAS fragmentation function measurements at $\sqrt{s}=7$~\TeV~\cite{ATLASfrag} and is thus expected
to describe inclusive-jet properties well.

Additional pile-up events, which are superimposed on the hard-scattering event, are generated with either
{\sc Pythia} 6~\cite{pythia} with the {\sc AUET2B} tune using the MRST LO** PDF~\cite{Martin:2009iq}
set, or {\sc Pythia} 8~\cite{Sjostrand:2007gs} with the {\sc 4C} tune~\cite{4CTune} using the CTEQ 6L1 PDF set~\cite{Pumplin:2002vw}.
Choosing between these two pile-up simulations has negligible impact on the analysis. The
number of pile-up events in the MC simulation is reweighted to match the conditions
found in the data for each trigger selection.  The events are passed through the ATLAS detector
simulation~\cite{simulation}, based on {\sc GEANT}4\cite{geant}, and are reconstructed using the same
software as for the data.

\subsection{Jet reconstruction, selection and calibration}
Jets are constructed from topological clusters of calorimeter cells~\cite{Lampl:1099735} and
calibrated using the EM+JES scheme~\cite{jesPaper,InclJets}.
This scheme is designed to adjust the energy measured in the calorimeter to that of the
true particle jets on average.
Calorimeter jets are reconstructed using
the \AKT jet algorithm~\cite{Cacciari:2008gp,FastJet} with a four-momentum recombination scheme
and studied if calibrated transverse momentum $\pt>20$~\GeV\ and $|\eta|<4.5$. 
Jet-finding radius parameters of both $R=0.4$ and $R=0.6$ are studied. Only jets with 
$|\eta|<2.1$ are used for building the quark-jet tagger, to guarantee that the jet is well within
the tracking acceptance. 
In the MC simulation, particle jets are reconstructed using the same \AKT algorithm with stable, interacting particles~\footnote{
A particle is considered stable and interacting if its lifetime is longer than 10~ps and it is neither a muon nor a neutrino.
} as input to the jet algorithm.  
In all cases, jet finding is done in $({\rm rapidity},\phi)$ coordinates and jet calibration is done in $(\etajet,\phijet)$ coordinates.

The reconstructed jets are additionally required to satisfy several data quality and isolation criteria.
The data quality cuts are each designed to mitigate the impact of specific non-collision backgrounds~\cite{InclJets}.
Reconstructed and particle jets are considered isolated if there is no other reconstructed jet
(or particle jet) within a cone of size $\DeltaR=\sqrt{(\Delta\eta)^2+(\Delta\phi)^2}<0.7$ around the jet axis.
Only isolated jets are considered in this study. 
The jet vertex fraction (JVF) is calculated for each jet and used to reject 
jets originating from pile-up interactions. The JVF is built using information about the origin, 
along the direction of the beam, of tracks with $\DeltaR<0.4$ ($\DeltaR<0.6$) to the jet axis 
for $R=0.4$ ($R=0.6$) jets and describes the fraction of the jet's
charged particle $\pt$ associated with the primary vertex~\cite{jesPaper}.

\subsection{Track selection and associating tracks with jets}\label{sec:trackSel}

Tracks are associated with jets by requiring that the track momentum direction (calculated
at the primary vertex) and the jet direction satisfy $\DeltaR({\rm jet,track})<0.4$
($\DeltaR({\rm jet,track})<0.6$) for $R=0.4$ ($R=0.6$) jets.  
Track parameters are evaluated at the point of closest approach to the primary 
hard-scattering vertex, which is the vertex with the highest sum of associated track $\pt^2$.
Tracks are required to have $\pt>1$~\GeV, at least one pixel hit and at least six hits
in the silicon strip tracker, as well as transverse (longitudinal) impact parameters with respect
to the hard-scattering vertex $|d_0|<1$~mm ($|z_0\cdot \sin(\theta)|<1$~mm).

The studies in this paper were also performed with a requirement of track $\pt>500$~\MeV.  No
significant changes to the results were found.  Requiring $\pt>1$~\GeV\ reduces 
the sensitivity to pile-up and the underlying event, and this requirement is used for the
remainder of the paper.  A ``ghost association''~\cite{ghost} procedure  
was also tested in place of $\DeltaR$-based matching, and no significant differences are observed. 
The jet isolation requirement helps to guarantee the similarity of the ghost association procedure and the $\DeltaR$ matching.

\subsection{Photon selection}

Photons with $\pt>25$~\GeV\ are selected with pseudorapidity $|\eta|<2.37$, 
excluding the transition region between the barrel and end-cap calorimeters ($1.37<|\eta|<1.52$).
Only the leading photon in the event is considered. The photons are required to satisfy the preselection
and ``tight'' photon cuts described in Ref.~\cite{promptPhot}. An additional isolation cut requiring
less than $5\GeV$ of transverse energy in a cone of size $\DeltaR=0.4$ around the photon
is imposed to increase the purity of the sample~\cite{jesPaper}. The photons are additionally
required to be well separated from calorimeter defects and to not be within $\DeltaR<0.4$ of a jet arising
from non-collision backgrounds or out-of-time pile-up.

\subsection{Lepton selection}

Isolated electrons and muons are used to select $W+$jet and \ttbar\ events.
Electron candidates are formed by matching clusters found in the electromagnetic calorimeter to
tracks reconstructed in the ID in the region $|\eta|< 2.47$ and are required to have
transverse energy $\ET > 25 \GeV$. To ensure good containment of electromagnetic showers in the calorimeter,
the transition region $1.37 <|\eta| < 1.52$ is excluded as for photons.
The electron candidates must pass the ``tight'' selection criteria based on the lateral and transverse
shapes of the clusters described in Ref.~\cite{egammaPaper} but updated for 2011 running conditions.
Reconstructed tracks in the ID and the MS are combined to form muon candidates, which are selected
in the region $|\eta|<2.5$ and are required to have $\pt > 20 \GeV$.
The selection efficiency for electrons and muons in simulated events, as well as their
energy and momentum scale and resolution, are adjusted to reproduce those observed in
$Z\rightarrow\ell\ell$ events in data~\cite{egammaPaper}. To reduce the contamination from
jets identified as leptons, requirements are placed on the total momentum carried by tracks within
$\DeltaR=0.3$ of the lepton and on calorimeter energy deposits within $\DeltaR=0.2$, excluding the
track and energy of the lepton itself.
For muons, the scalar sum of the $\pt$ of these neighbouring tracks must be less than $2.5 \GeV$, while the
sum of this close-by calorimeter $\ET$ must be less than $4 \GeV$. For electrons, the sum of
calorimeter $\ET$ must be less than $6 \GeV$. Additionally, leptons are required to be
consistent with originating from the primary hard-scattering vertex.
They are required to have $|z_0|<10$~mm, and
the ratio of $d_0$ to its uncertainty ($d_0$ significance) must be smaller than 3.0 for muons
and 10.0 for electrons, due to the wider distribution found in signal electrons caused by
bremsstrahlung.

\subsection{Trigger and event selection}

All events must have a vertex with at least three associated tracks with $\pt>150$~\MeV.
Other event selection requirements are described below.

\subsubsection{Dijet and trijet samples}

The dijet sample is selected using single-jet triggers with various thresholds~\cite{triggerPaper},
which are fully efficient for jets with $\pt>40$~\GeV. Each jet $\pt$ bin is filled exclusively by a single
trigger that is fully efficient for jets in that $\pt$ range, following Ref.~\cite{InclJets}.
The trijet sample uses the same trigger selection as the dijet sample. This guarantees
that studies using the jet with the third highest $\pt$ in each event are not biased by the trigger.

\subsubsection{$\gamma$+jet and $\gamma$+2-jet samples}

The \gjet\ sample is selected using single-photon triggers.  The lowest threshold single-photon trigger
is fully efficient for photons with $\pt>25$~\GeV. For this sample, a back-to-back
requirement for the photon and the leading jet, $\Delta\phi>2.8$, is imposed. An additional
veto on soft radiation is also applied to further reduce background contamination~\cite{jesPaper}:
the uncalibrated  $\pt$ of the sub-leading jet is required to be
less than 30\% of the photon $\pt$. Relying on the $\pt$ balance of the photon and jet, each jet $\pt$ bin
is filled exclusively by a single-photon trigger that provides a fully efficient selection.

The same triggers are used in the $\gamma$+2-jet sample in each region of jet $\pt$. Since the
sub-leading jet $\pt$ is lower than that of the leading jet by definition, this selection is also not
biased by jet reconstruction effects.

\subsubsection{$W+$jet sample}

The $W+$jet sample is selected using a single-electron or single-muon trigger.
The event selection, following
Ref.~\cite{Wnote}, requires exactly one charged lepton (electron or muon) and that it matches the
trigger accepting the event, a transverse mass~\footnote{$\mt = \sqrt{\MET \times \ET^\ell \times (1-\cos(\Delta\phi))}$,
where \MET\ is the missing transverse momentum in the event, $\ET^\ell$ is the lepton transverse energy
(transverse momentum for a muon), and $\Delta\phi$ is the angle between the lepton and the \MET\ in the $\phi$ direction.}
$\mt> 40\GeV$, missing transverse momentum $\MET>25\GeV$, and at most two jets (to reject $\ttbar$ backgrounds).
The triggers are fully efficient for electrons and muons satisfying the offline $\pt$ requirements.

In events in which two jets are reconstructed, only the jet with the highest $\pt$ is studied.

\subsubsection{$\ttbar$ sample}

Top quark pair events in which exactly one of the $\Wboson$ bosons produced by the top quarks decays to
an electron or a muon are selected as described in Ref.~\cite{mtop2011paper}. The event selection requires
that exactly one electron or muon is reconstructed and that it matches the trigger accepting the event.  Background
suppression cuts of $\mt>40\GeV$ ($\mt>60\GeV$) and $\MET>25\GeV$
($\MET>20\GeV$) in the electron (muon) channel, and at least four jets with $\ptjet> 25\GeV$, $|$JVF$|>0.75$
and $|\etajet|<2.5$ are also required.
Two of the selected jets must be identified
as arising from a $b$-quark ($b$-tagged) using the
MV1 algorithm, which combines several tracking variables into a multi-variate discriminant,
with the 60\% efficiency working point~\cite{MV1}.

After this selection, the background contamination in the $\ttbar$ sample is of the order of 10\% and
consists mainly of events from $W/Z$+jets or single top-quark production. The contribution from
multi-jet background after the requirement of two $b$-tagged jets is about $4\%$~\cite{mtop2011paper}.
The background contamination in the selected data sample has no sizable impact in the studies
performed. The change in the results when including the background in the analysis is small, and the
sample is therefore assumed to be pure $\ttbar$.

\section{Jet labelling in Monte Carlo simulation}
\label{sec:jetdef}


One natural definition of the partonic flavour of a jet in a Monte Carlo event is given by matching the
jet to the closest outgoing parton (in $\DeltaR$) from the matrix-element calculation, which
represents a fixed-order QCD event record. In generators with $2\rightarrow 2$ matrix elements,
such a matching scheme is clear only for the two leading jets at most.
To simplify the task for analyses using different MC simulations, jets are matched to
the highest-energy parton in the parton shower record within a $\DeltaR$ equal to the radius
parameter of the jet algorithm. Using this method, only a small fraction of the jets ($<1$\% around
jet $\pt=50$~\GeV\ and fewer above 100 \GeV) are not assigned a partonic flavour. Studies with {\sc Pythia} 6
and {\sc MadGraph} indicate that jets with significant energy contributions from more than one
distinct parton (e.g. overlap of initial- and final-state radiation) are rare in
the samples used.
The jet isolation requirement restricts the wide-angle QCD radiation of the jet and further guarantees the accuracy
of the labelling based on the parton shower record.

Jets are identified as originating from $c$- and $b$-quarks by requiring one $c$- or $b$-hadron with
$\pt>5$~\GeV\ in the MC record within a $\DeltaR$ equal to the radius parameter of the jet.
Jets with two $c$- or $b$-hadrons are identified
as including a gluon splitting to $\ccbar$ or $\bbbar$.  Both classes are considered separately from
quark- and gluon-jets. The labelling of $b$-jets supersedes that of $c$-jets, which itself supersedes the
quark and gluon labelling. In the samples used, other than $\ttbar$, the fraction of heavy-flavour jets is relatively small.
The variables used for quark- and gluon-jet discrimination are sufficiently different
for each of these jet types to require an independent treatment.

In MC event generators with matching schemes~\cite{MLM,CKKW,CKKWL}, it is possible to use the outgoing
partons from the matrix-element calculation to label jets. Only jets above the matching
scale can be identified in this manner, and only in exclusively showered events (\ie\ events with the
same number of jets at the matrix-element level and after showering).  To avoid the need to tag jets
originating from partons created in the parton shower, the matching scale must be chosen to be
much lower than the minimum $\pt$ of the jets for which the tagger is designed and commissioned.
Labelling of jets based on the highest-energy parton is consistent with labelling based on the
matrix-element calculation for isolated jets in the samples used here.  The former is therefore
used in this paper.

For the construction of templates and the examination of data, only ensembles of jets are considered.
The parton record of the MC simulation is not used. Instead, the fractions of quark- and
gluon-jets in each sample are calculated using the matrix-element event record, and only these fractions are
used to describe the average composition of the jet ensemble.

\section{Determination of quark-jet and gluon-jet properties}
\label{sec:mcTemp}


In previous theoretical~\cite{schwartz1} and experimental~\cite{jesPaper} studies, the jet width and
the number of tracks associated with the jet were found to be useful for identifying the partonic origin
of a jet. As discussed in Sec.~\ref{sec:intro}, the larger colour factor associated with a gluon
results in the production of a larger number of particles
and a softer hadron $\pt$ spectrum after the shower. To define the optimal discriminant,
several jet properties are examined for their ability to distinguish the partonic origin of a jet
and for their stability against various experimental effects, including pile-up.
As these jet properties depend on the jet kinematics, the analysis of the properties and the
resulting discriminant are separated into bins of jet $\pt$ and $\eta$.  The $\pt$ bin width is dictated
by a combination of the jet resolution and the number of available events in data, and the $\eta$ bins 
coarsely follow the detector features.

\subsection{Discriminating variables}
\label{sec:variables}

Useful discriminating variables, such as the number of particles associated with a jet, may be
estimated using either the number of charged-particle tracks in the inner detector or using
the number of topological clusters of energy inside the jet~\cite{jesPaper}.
Although they are
limited to charged particles, and thus miss almost half of the information in a typical
jet, jet properties built from tracks have three practical advantages over calorimeter-based properties.
First, they may include particles that have sufficiently low $\pt$ that they are not measured by
the calorimeter, or which are in the regime where the ID momentum measurement is
more accurate than the energy measurement of the calorimeter. Second, charged
particles bend in the magnetic field of the ID. Additional particles from the
underlying event brought into the jet produce a background in the calorimeter, and
particles that are sufficiently bent are lost to the calorimeter jet. However, both classes of particles
can be correctly assigned using their momenta calculated at the interaction point.
Third, tracks can be easily associated with a specific vertex. This association dramatically
reduces the pile-up dependence of track-based observables. Similar arguments hold in the
calculation of jet shape variables.

The variables surveyed as potential inputs to the quark/gluon tagging discriminant are:

\begin{itemize}
\item Number of reconstructed tracks ($n_{\rm trk}$) in the jet.
\item Calorimeter width:
\begin{equation*}
w = \frac{ \sum_i p_{{\rm T},i} \times \Delta R(i,{\rm jet})}{ \sum_i p_{{\rm T},i} }, 
\end{equation*}
where the sum runs over the calorimeter energy clusters that are part of the jet.
\item Track width, defined similarly to the calorimeter width but with the sum running over associated tracks.
\item Track-based energy-energy-correlation (EEC) angularity:
\begin{equation*}
{\rm ang_{\rm EEC}}=\frac{ \sum_i \sum_j p_{{\rm T},i} \times p_{{\rm T},j} \times (\Delta R(i,j))^\beta}{ (\sum_i p_{{\rm T},i})^2 },
\end{equation*}
where the index $i$ runs over tracks associated with the jet, $j$ runs over tracks associated with the jet while $j>i$,
and $\beta$ is a tunable parameter~\cite{OriginalEEC,Larkoski2013eya}.
\end{itemize}

\begin{figure}[tbp]
\begin{center}
\includegraphics[width=0.45\textwidth]{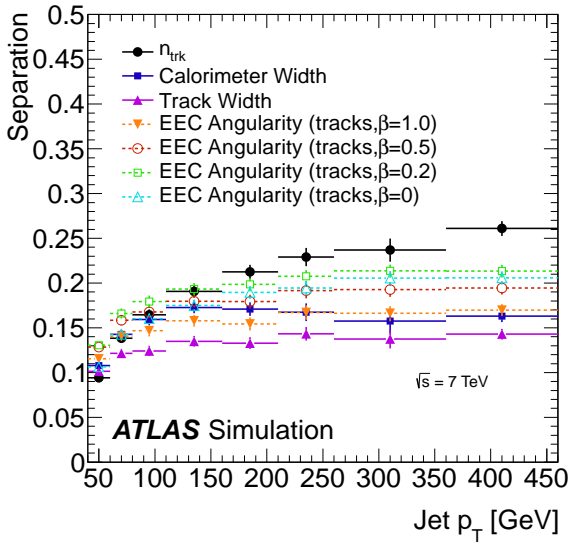}
\caption{Separation power provided by different variables between quark- and gluon-jets as
a function of jet $\pt$ in the {\sc Pythia} 6 dijet MC simulation for jets with $|\eta|<0.8$ built
with the anti-$k_t$ algorithm with $R=0.4$.  }
\label{fig:someDiscriminants}
\end{center}
\end{figure}

The discriminating power (``separation'') of a variable $x$ is calculated
as in Ref.~\cite{2006physics11219P} to investigate the effectiveness of each variable in a quark/gluon tagger in a
sample with equal fractions of quarks and gluons:

\begin{equation*}
s=\frac{1}{2}\int\frac{(p_q(x)-p_g(x))^2}{p_q(x)+p_g(x)}dx = \frac{1}{2}\sum_i\frac{(p_{q,i}-p_{g,i})^2}{p_{q,i}+p_{g,i}}, 
\label{eq:separation}
\end{equation*}

\noindent where $p_q(x)$ and $p_g(x)$ are the normalised distributions of the variables for quark- and gluon-jets,
and where the second expression applies to histograms, with the sum running over the bins of the histogram.
This definition corresponds to the square of the statistical uncertainty that one would
get in a maximum-likelihood fit when fitting for the fraction of quark- or gluon-jets using the
given variable, divided by the square of the uncertainty in the case of perfect separation.
While this is not a variable that relates easily to quantities of interest for tagging, its interpretation
is independent of the shape of the distributions, allowing for comparisons that are independent
of the tagging efficiency. 
Using this definition, Fig.~\ref{fig:someDiscriminants} shows, for different variables, the separation
between quark-jets and gluon-jets as a function of jet $\pt$ for jets built with
the anti-$k_t$ algorithm with $R=0.4$ using the {\sc Pythia} 6 dijet MC simulation.
In this simulation, the two most powerful variables are the EEC angularity with $\beta=0.2$ and the
number of tracks associated with the jet. 
The jet width built using the associated tracks is the weakest discriminant and the calorimeter-based width is
somewhat stronger, and of comparable power to that of the EEC angularity with $\beta=1.0$.

All track-based variables show excellent stability against pile-up and 
significant discrimination power between quark- and gluon-jets. 
The dependence of the mean calorimeter width on the number of reconstructed vertices is about five times stronger 
than the dependence of any of the variables considered for the final discriminant and at low jet $\pt$ is 
up to $\approx 1.5$\% per primary vertex.  At high jet $\pt$, the dependence is negligible for all variables.
While it is possible to correct the
inputs or to use a pile-up-dependent selection to allow the use of calorimeter-based  
variables without introducing a pile-up dependence in the tagger, such an approach is not pursued in this paper.  
Although Fig.~\ref{fig:someDiscriminants} suggests using the charged particle multiplicity and the EEC angularity with
$\beta=0.2$ to build the tagger, a larger linear correlation between these two variables makes this
tagger perform worse at high $\pt$ than the tagger built using the charged particle multiplicity
and the track width. Furthermore, differences between data and MC simulation are
reduced when using the latter tagger. For this reason, track width and $n_{\rm trk}$ are
used to build the discriminant used in the rest of this paper.
The linear correlations between $n_{\rm trk}$ and track width  
are at the 15\% level at low $\pt$, increasing to 50\% at high $\pt$. Thus, the variables add
independent information about the properties of the jet. 
For EEC angularity with $\beta=0.2$, the linear correlation with $n_{\rm trk}$ is about 75\% with a weak dependence
on $\pt$.
Still, the study of the EEC angularities and the evolution of their tagging performance as a function of
$\beta$ is interesting for reasons discussed in Ref.~\cite{Larkoski2013eya}. Since this discussion
is not relevant for the tagger developed in this paper, it is relegated to \ref{sec:EECresults}.

\subsection{Extraction of pure templates in data}\label{sec:extraction}

To construct a discriminant, the properties of ``pure'' quark- and gluon-jets must be determined.
As these properties depend on the modelling of non-perturbative effects, they are extracted from data
to avoid reliance on MC simulations. The extraction
can be performed using unbiased samples of pure quark- and gluon-jets or, alternatively,
several mixed samples for which the admixture is well known theoretically.
The use of pure samples is explored in detail in Sec.~\ref{sec:disc} as a 
validation procedure but is not used to determine the performance of the tagger in data, due
to the limited number of events available and the difficulties in obtaining samples with negligible
gluon and light-quark contaminations. The use of mixed samples is described below in detail, since it is
used to create an operational tagger for data.

Distributions of properties of quark-jets or gluon-jets are extracted using the dijet and
$\gamma+$jet event samples and the fraction of quark- and gluon-jets predicted by
{\sc Pythia} 6 with the {\sc AUET2B} tune. 
For each bin $i$ of jet $\eta$, jet $\pt$, and jet property (track width, number of tracks, or the two-dimensional
distribution of the these), a set of linear equations is solved:

\begin{align}
P_i(\eta,\pt) &= f_{q}(\eta,\pt) \times P_{q,i}(\eta,\pt) \nonumber \\
    & + f_{g}(\eta,\pt) \times P_{g,i}(\eta,\pt) \nonumber \\
& + f_{c}(\eta,\pt) \times P_{c,i}(\eta,\pt) \nonumber \\
& + f_{b}(\eta,\pt) \times P_{b,i}(\eta,\pt) , 
\label{eq:fractions}
\end{align}

\noindent where $P_i$ is the value of the relevant distribution in bin $i$ of the distribution in the dijet
or $\gamma+$jet sample, $f_{q}$ and $f_{g}$ are the light-quark and gluon fractions
predicted by {\sc Pythia} at a given $\eta$ and $\pt$, 
and $P_{q,i}$ and $P_{g,i}$ are the values of the relevant distribution
for quark- and gluon-jets in bin $i$ of the distribution. The fractions  
$f_{c}$ and $f_{b}$ for $c$-jets and $b$-jets are relatively small. They are taken from the
MC simulation, together with the corresponding distributions $P_{c}$ and $P_{b}$.
The same is true for the fractions and distributions for $g\rightarrow c\bar{c}$ and $g\rightarrow b\bar{b}$,
not shown in Eq.~\ref{eq:fractions} for brevity. 
By using the different fractions of light quarks and gluons in dijet and $\gamma+$jet events in
each $\pt$ and $\eta$ bin, the expected ``pure'' jet sample properties ($P_{q}$ and $P_{g}$) 
can be estimated. In these samples, the $b$-jet and $c$-jet fractions are typically below $5-10\%$. 
The studies are performed in three bins of $|\eta|$: $|\eta|<0.8$, $0.8<|\eta|<1.2$ and $1.2<|\eta|<2.1$.

An additional term $f_{{\rm fake}, i}(\eta,\pt)\times P_{{\rm fake}, i}(\eta,\pt)$
must be added to the 
distributions in the $\gamma+$jet sample to account for events in which the reconstructed  
(``fake'') photon arises from a jet with energy deposits mostly within the electromagnetic calorimeter.
The term is estimated from data using a sideband counting technique, developed and implemented in
Refs.~\cite{jesPaper} and~\cite{promptPhot}. The method uses regions defined with 
varying levels of photon isolation and photon identification criteria, estimating the number
of background events in the signal region from those in the background regions, after
accounting for signal leakage into the background regions.

\begin{figure*}[tp]
\begin{center}
\subfigure[]{
\includegraphics[width=0.45\textwidth]{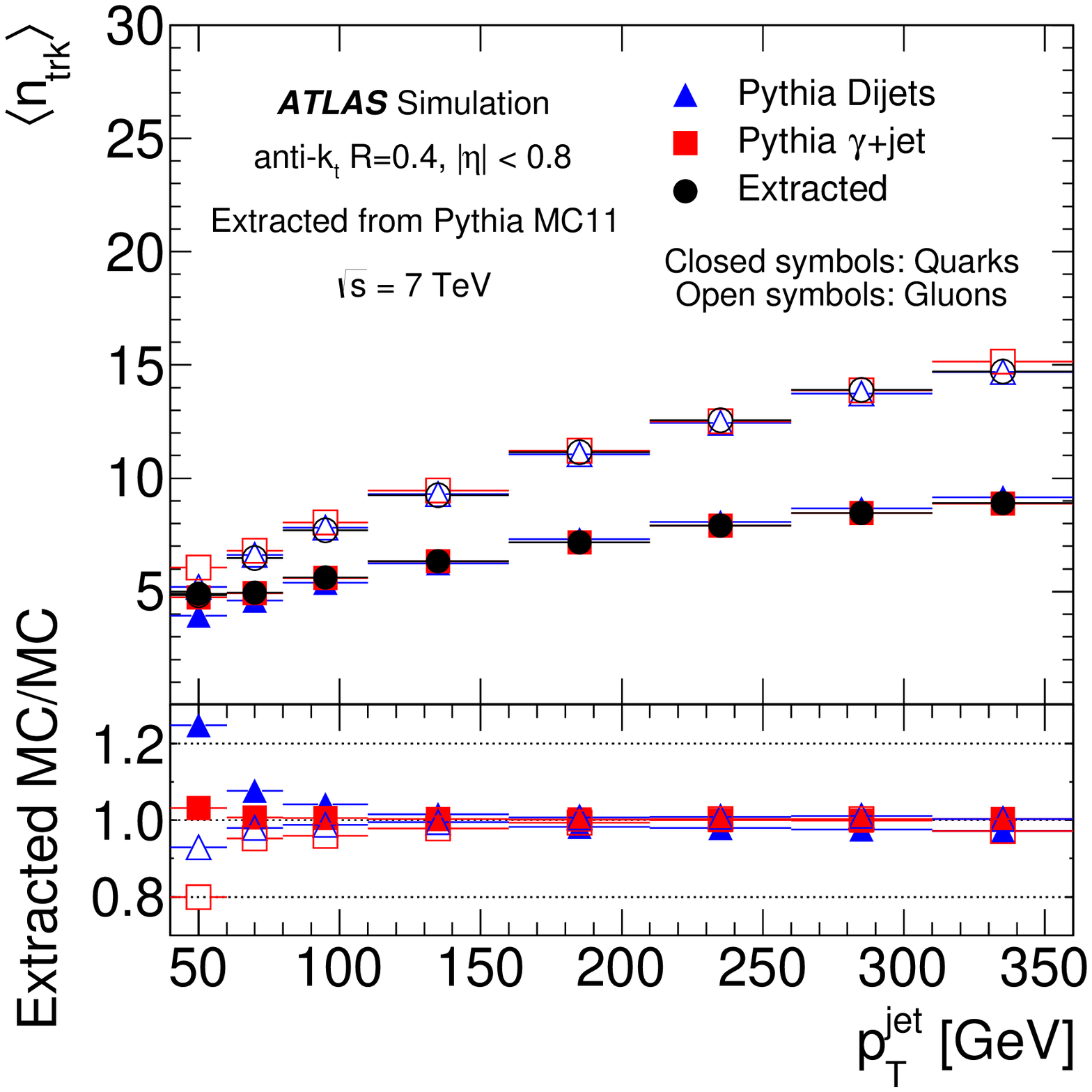}
}
\subfigure[]{
\includegraphics[width=0.45\textwidth]{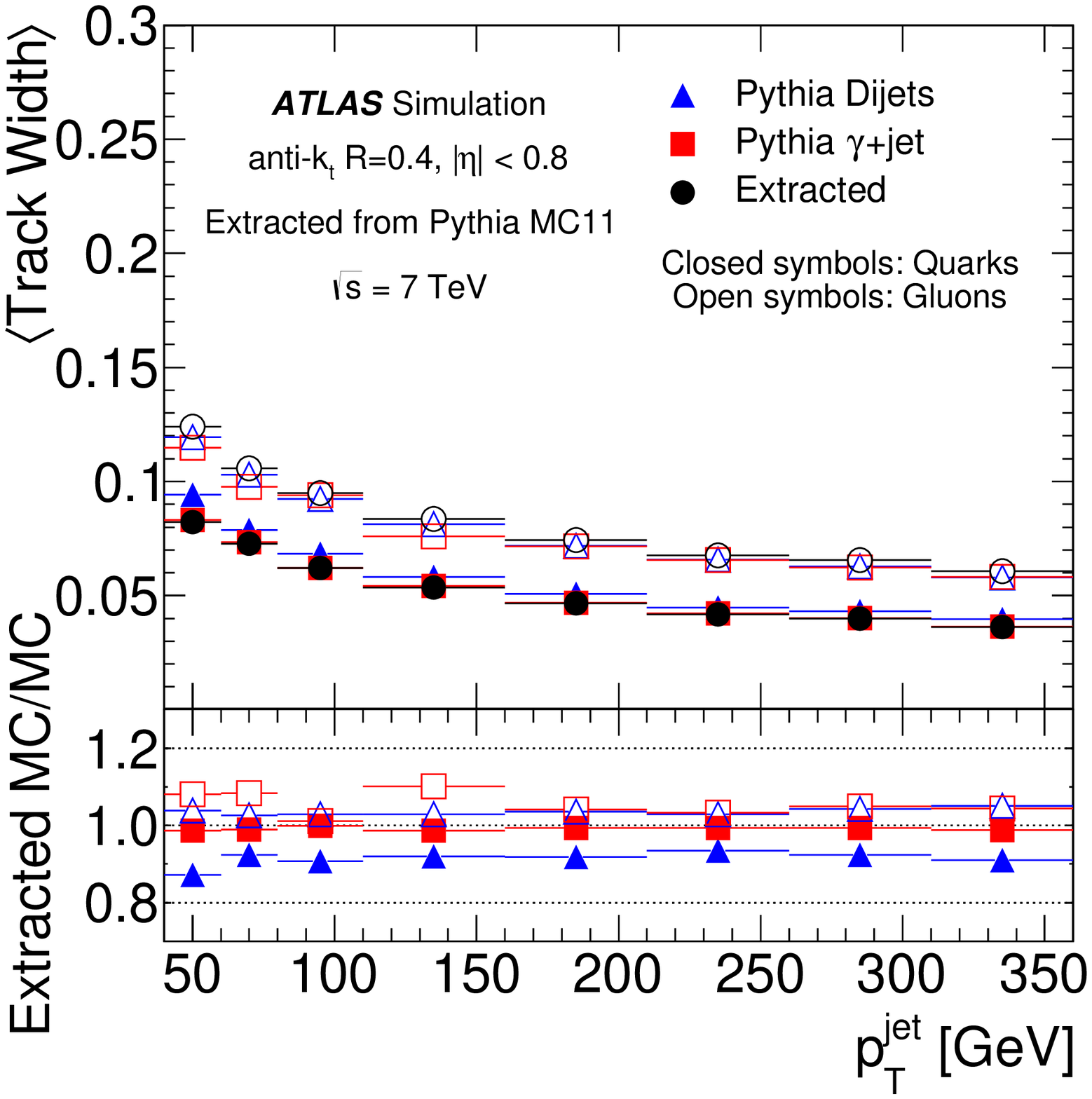}
} \\
\subfigure[]{
\includegraphics[width=0.45\textwidth]{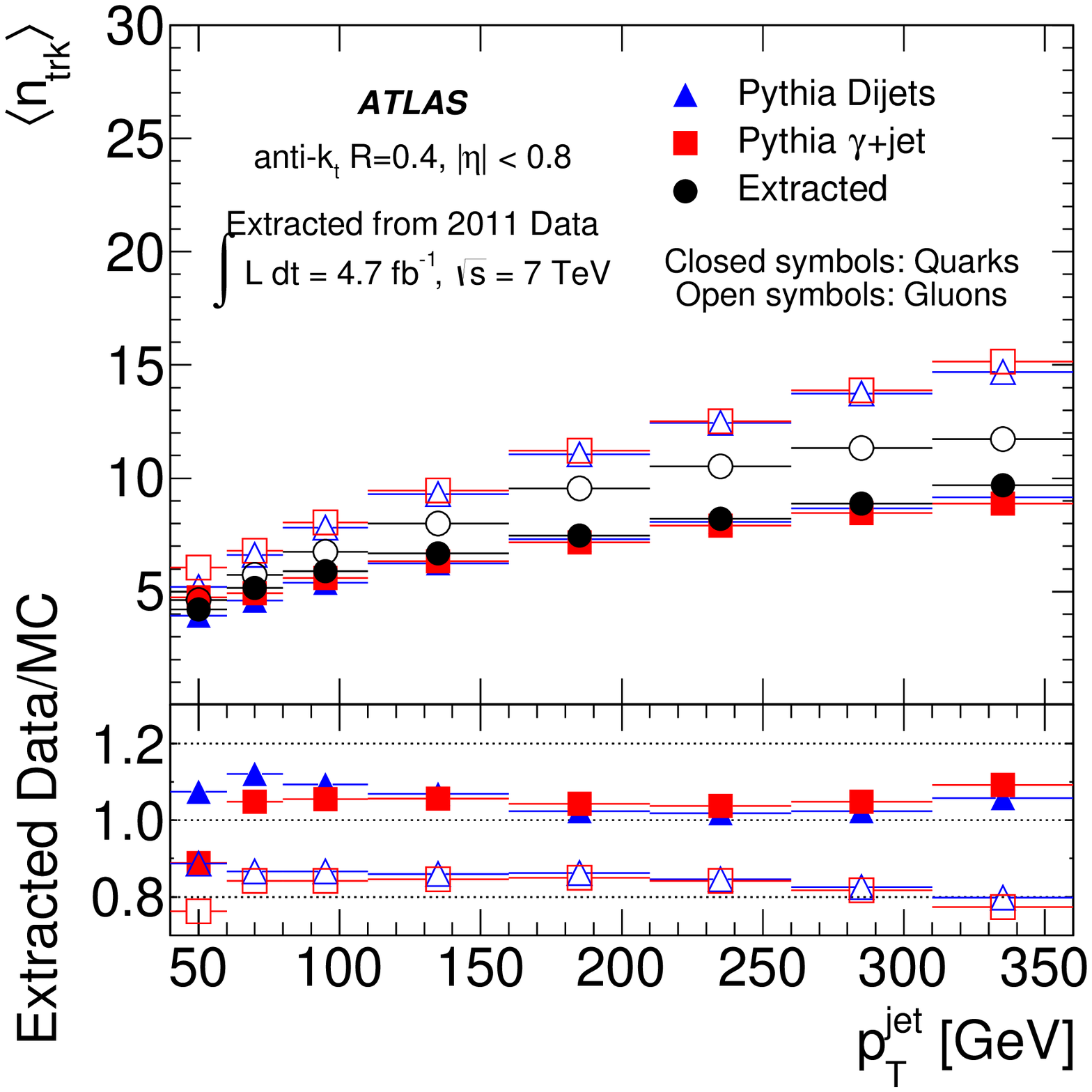}
}
\subfigure[]{
\includegraphics[width=0.45\textwidth]{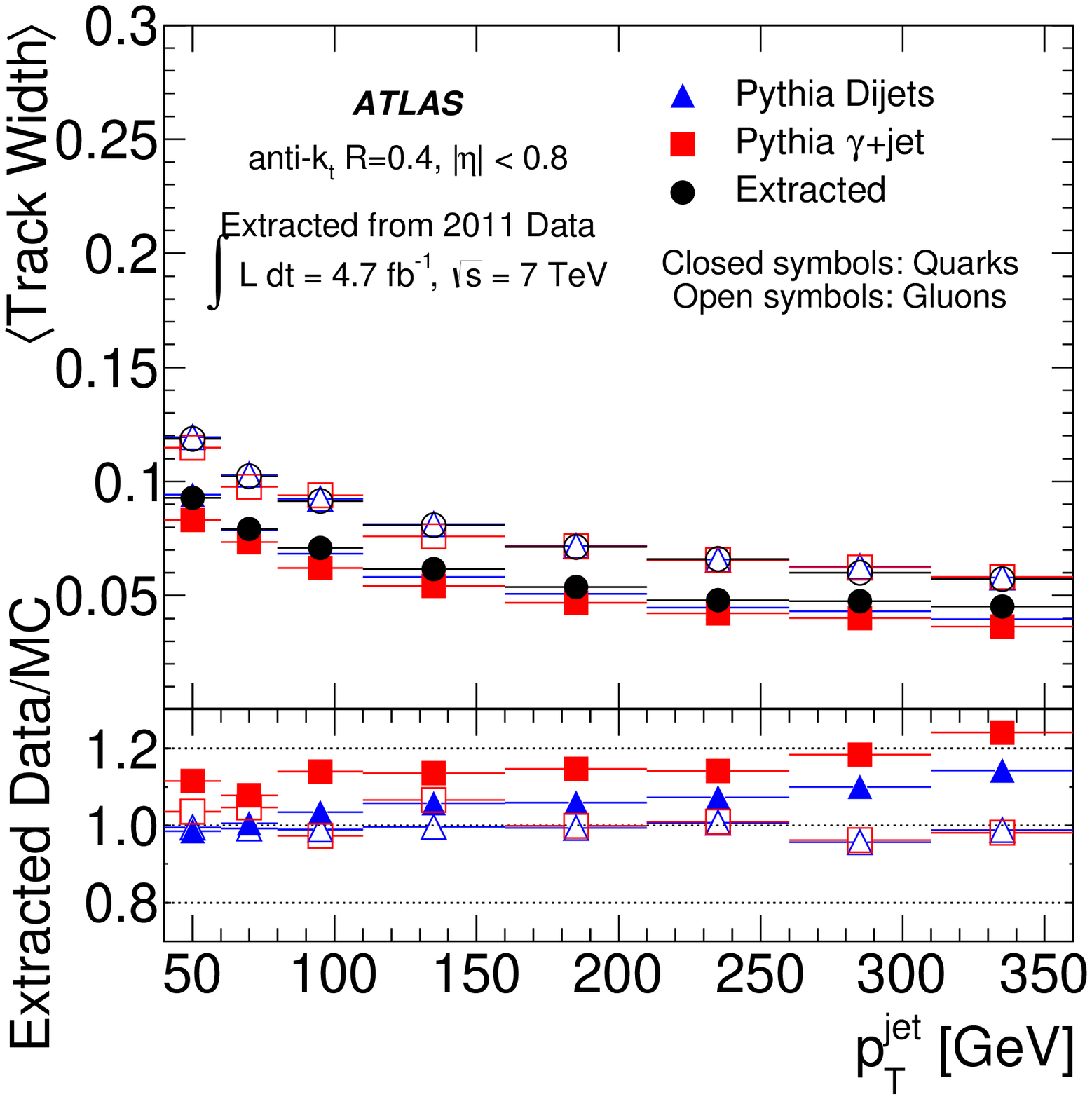}
}
\caption{ Average (a,c) $n_{\rm trk}$ and (b,d) track width for quark- (solid symbols) and gluon-jets
(open symbols) as a function of reconstructed jet $\pt$ for isolated jets with $|\eta|<0.8$.
Results are shown for distributions obtained using the in-situ extraction method in 
{\sc Pythia} 6 simulation (black circles, (a,b)) or data (black circles, (c,d)), as well as for
labelled jets in the dijet sample (triangles) and in the $\gjet$ sample (squares).
The error bars represent only statistical uncertainties. Isolated jets are reconstructed using the
anti-$k_t$ jet algorithm with radius parameter $R=0.4$. The bottom panels show
the ratio of the results obtained with the in-situ extraction method to the results
in the dijet and $\gjet$ MC samples. }
\label{fig:ntrkWidthClos}
\end{center}
\end{figure*}

Knowledge of $P_x$ and $f_x$ for the dijet and $\gamma+$jet samples allows the
extraction of pure quark- and gluon-jet $n_{\rm trk}$ and track width distributions from the
data. The method can be tested in the MC simulation, comparing the properties
of jets labelled in MC as quark- or gluon-jets and the properties extracted using Eq.~(\ref{eq:fractions})
to demonstrate consistency. Figure~\ref{fig:ntrkWidthClos} (top) shows the mean number of tracks and the mean track width
as a function of the jet $\pt$, separated using either the MC flavour labels 
or the extraction procedure in the same MC events for jets with $|\eta|<0.8$.  
Differences are observed between the average of the distributions in the dijet and 
$\gamma+$jet samples. This biases the extracted distributions for gluon-jets to be more
like the gluon-jet properties in the dijet sample.  The same is true for quark-jets and
the $\gamma+$jet sample.  The 
differences are larger at low $\pt$ and for the track width distributions. The bias demonstrates
a sample dependence, which is included as a systematic uncertainty on
the performance of the discriminant built from these jet properties. 
These differences are, however, small compared to the differences between quark- and
gluon-jets, demonstrating the sensitivity of the extraction method. Similar results
are obtained for jets reconstructed with radius parameter $R=0.6$ and in other $|\eta|$ regions.

Figure~\ref{fig:ntrkWidthClos} (bottom) shows the same MC simulation points as Fig.~\ref{fig:ntrkWidthClos} (top), but
here the data are used in the extraction.  Relatively good agreement is found between data and {\sc Pythia}
{\sc AUET2B} for the track width of gluon-jets and for the number of tracks in quark-jets. 
However, the mean number of associated tracks is significantly smaller for gluon-jets in the
data than in the {\sc Pythia} MC simulation. Similarly, the mean 
track width is larger in data than in the MC simulation for quark-jets.

Both these differences make gluon-jets and quark-jets more similar, reducing the discrimination
power of these properties in data. Differences between the {\sc Pythia} MC 
simulation and the data are also present in some of the other variables originally considered. 
These differences translate into non-negligible differences in the corresponding discriminants.
For this reason, a fully data-driven tagger is built.

\subsection{Systematic uncertainties on the extraction procedure}\label{sec:systematicsExtr}

The distributions extracted from data can be used to build a data-driven tagger, and to evaluate its
performance in data. 
Uncertainties on the extracted pure quark- and gluon-jet property templates are thus propagated
through as uncertainties on the performance of the tagger. The systematic effects considered can
be classified into four categories:  
uncertainties on the input fractions ($f_{x,i}$),
uncertainties on the input shapes ($P_{x,i}$),
uncertainties on the fake photon background,
and sample-dependence effects.
This last category includes, for example, differences in quark-jet properties between samples, 
which result in different quark-jet rejection across the various samples. This effect is the one
that causes the inconsistency in the extraction method, illustrated in Fig.~\ref{fig:ntrkWidthClos}. 
Sample-dependent effects are included as a systematic uncertainty rather than deriving a separate tagger for
each event selection and MC simulation.

Because jets with different observable properties have different calorimeter response, an  
additional uncertainty in the jet energy scale arises from the modelling of the response as a function of the discriminant
in the MC simulation.  The resulting uncertainties on the jet energy response after tagging, in addition to the standard 
jet energy scale uncertainties,  are determined to be below 1\% using a $\gjet$ $\pt$-balance study following the procedures
described in Ref.~\cite{jesPaper}.

\subsubsection{Input fraction uncertainties}

The fraction of quark- and gluon-jets can change 
when going from a leading-order calculation to a next-to-leading-order (NLO) calculation, changing
the renormalisation/factorisation scale, or changing the PDF set.

The first two effects are examined by comparing the {\sc Pythia} and {\sc MadGraph} 
calculations, which have different renormalisation/factorisation scales and different ways  
of simulating real emissions. Similarly, the potential effect of the real emissions is also probed by
comparing the matrix-element labelling and the highest-energy parton labelling. A 5\% uncertainty
that is anti-correlated between quark- and gluon-jets 
is applied to cover the maximum variation seen in these comparisons. This uncertainty
is uncorrelated amongst samples.

The potential mis-modelling of the fraction of quark- and gluon-jets in the MC
simulation due to limitations of the PDFs is estimated using several 
PDF sets.  The PDF sets use different fitting procedures (MRST, CTEQ and NNPDF sets), 
different orders in the perturbation theory expansion (MSTW2008lo for LO, CT10 for NLO) 
and different assumptions about the $\alpha_s$ calculation (MRST2007lomod for LO$^*$ and
MRSTMCal for LO$^{**}$). A 5\% uncertainty, anti-correlated between quark- and gluon-jets, conservatively covers the
differences between the various PDF sets. This uncertainty is considered uncorrelated between 
the dijet and $\gjet$ samples because no significant trend is observed between the samples as the
PDF set is changed.

\subsubsection{Heavy-flavour input uncertainties}

The fractions of $b$-jets and $c$-jets are varied by $\pm20$\% in the dijet sample, following Ref.~\cite{HFfraction},
 and by $\pm50$\% in the $\gamma+$jet sample to estimate a conservative uncertainty. As
the fractions of $b$-jets and $c$-jets are small, these uncertainties remain sub-leading.
The two input fractions are varied independently. The differences in the results obtained 
after the extraction of the pure quark- and gluon-jet properties are added in quadrature
to obtain the total systematic uncertainty from this effect.

Uncertainties on the properties of $b$-jets are determined using a $\ttbar$ sample,
described in Sec.~\ref{sec:data-selection}. The purity of this sample is generally better than
95\%.  An envelope 10\% uncertainty is included on the $b$-jet properties as a result of
comparisons of $b$-jet properties between data and several MC simulations. The validation is
performed using tagged jets. Differences between tagged and inclusive $b$-jets in 
the MC simulation are found to be within the assigned uncertainty.

For $c$-jets, several templates with 10\% increases in the rates of 2-prong, 3-prong, and 4-prong
decays are used to estimate the effect of changes to the $c$-hadron decay.
These different $c$-jet distributions are propagated through the extraction procedure and the
largest difference is used as the systematic uncertainty on the performance of the 
tagger due to this effect. 

\subsubsection{Fake photon background uncertainty}

Several variations in the background to the $\gamma+$jet sample are considered.
The identification requirements used to define the regions for the background estimation method are
changed, resulting in purity differences of up to 10\% for low-$\pt$ jets. The same procedure 
is used to estimate an uncertainty on the jet properties in the fake background. An 
uncertainty of up to 4\% covers the changes in the means of the property distributions.
These differences are propagated to the discriminant distribution to obtain a systematic uncertainty
due to the purity estimate. An additional uncertainty covering the full shape correction to
$P_{\rm fake}$ for signal leakage into the background regions of the sideband counting method
is included as well, amounting to less than a 3\% change in the means of the property distributions.

\subsubsection{Sample-dependence uncertainty}

The application to a signal sample of a quark/gluon discriminant derived in a specific set of samples (or sample admixtures) rests
upon the assumption that sample dependence is negligible, or that it can at least be parameterised
as a function of visible properties of the event. One such property is the degree of isolation
of the jet, which requires separate treatment. However, there are other effects, such as colour flow, 
that are much harder to constrain using the available data and may lead to a sample-dependence of
jet properties.

Uncertainties on the jet properties are estimated first from differences
between the $\gamma+$jet and dijet samples of the properties of quark- and gluon-jets.  These are representative of
the differences observed when comparing several different samples.  Events generated
with {\sc Pythia} 6 and {\sc Herwig}++ are also tested for this effect.  The envelope of these
variations is used to estimate a systematic uncertainty due to the sample dependence
of the jet properties.
This systematic uncertainty is sensitive to statistical uncertainties in the MC simulation. These statistical
uncertainties
are estimated and used to smooth the $\pt$ dependence of the uncertainty following the
procedure described in Ref.~\cite{JESUpdate}. 
The sample dependence is consistently the dominant systematic uncertainty for all jet $\pt$ bins.
The differences between MC labelled samples derive from differences
in observable properties in the dijet and $\gjet$ samples.  It is thus critical to consider these effects when
estimating uncertainties on the tagging efficiency.

The properties of non-isolated jets differ from those of isolated jets, in general.
In both the data and the
MC simulation, isolated jet properties show no significant dependence on the $\DeltaR$ to
the nearest reconstructed jet for $\DeltaR>0.7$. As the discriminant constructed here uses only jets
satisfying this isolation criterion, no additional uncertainty due to the effect of jet
non-isolation is applied.

An additional uncertainty arises from an incorrect description of the $\pt$-dependence of
the tagging variables for samples with a significantly different jet $\pt$ spectrum from
that of the dijet and $\gamma+$jet samples with which the discriminant was constructed.
This accounts for the differences in bin-to-bin migrations in the various samples.
As this uncertainty is dependent entirely on the sample to which the discriminant is
applied, it is not explicitly included here.

\section{Validation with event-level kinematic cuts}
\label{sec:disc}


The jet property templates extracted in the previous section can be further
validated using high-purity quark- and gluon-jet samples.
Largely following the work in Ref.~\cite{schwartz2}, events are selected using basic kinematic cuts
and event-level selection criteria to study purified samples of quark-jets and gluon-jets.
These event selections are independent of the properties of individual jets and thus do
not bias them. By including several different selections, the importance of colour
flow and other sample-dependent effects can be evaluated using data.

The jets that are not tagged as $b$-jets in the $\ttbar$ sample, particularly in the case of
events with exactly four jets, are mostly light-flavour jets.  However, because of impurities
introduced by gluon contamination and $\Wboson\rightarrow c\sbar$ decays, they are not sufficiently
pure to be of use in this study.

\subsection{Validation of gluon-jet properties}\label{sec:gluonJets}

As protons have a large gluon component at low $x$, inclusive low-$\pt$ jet production at the LHC has a
high rate of gluon-jet production. However, the fractions drop rapidly as jet $\pt$ increases.
Particularly at moderate- and high-$|\eta|$, the relative rate of gluon-jet production
exceeds 50\,\% only below 150~\GeV\ in jet $\pt$.

Multi-jet events from QCD contain relatively more gluon radiation than the inclusive jet sample.
The radiation is typically soft, implying that the third-leading jet will
often be a gluon-jet.
A useful kinematic discriminant that can further purify a multi-jet sample, discussed
in Ref.~\cite{schwartz2}, is:

\begin{equation}
\zeta=|\eta_3|-|\eta_1-\eta_2|,
\end{equation}

\noindent where $\eta_i$ is the pseudorapidity of the $i$th leading jet. A selection based on
this variable can provide gluon-jet purity over 90\%, at the price of significantly reduced efficiency.

\begin{figure*}[tp]
\begin{center}
\subfigure[]{
\label{fig:pure_trijet_ntrkTrkWidth1}
\includegraphics[width=0.45\textwidth]{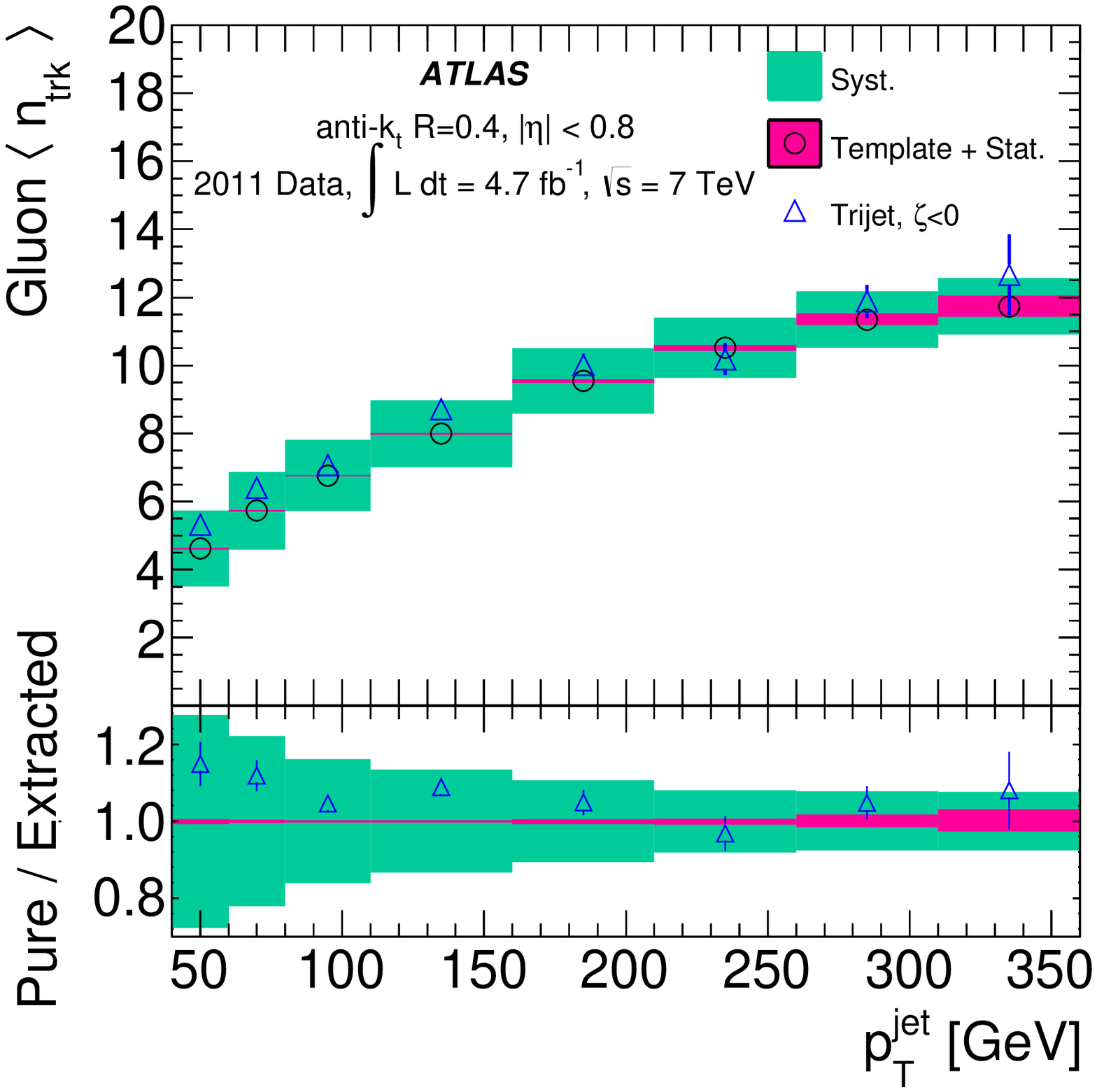}
}
\subfigure[]{
\label{fig:pure_trijet_ntrkTrkWidth2}
\includegraphics[width=0.45\textwidth]{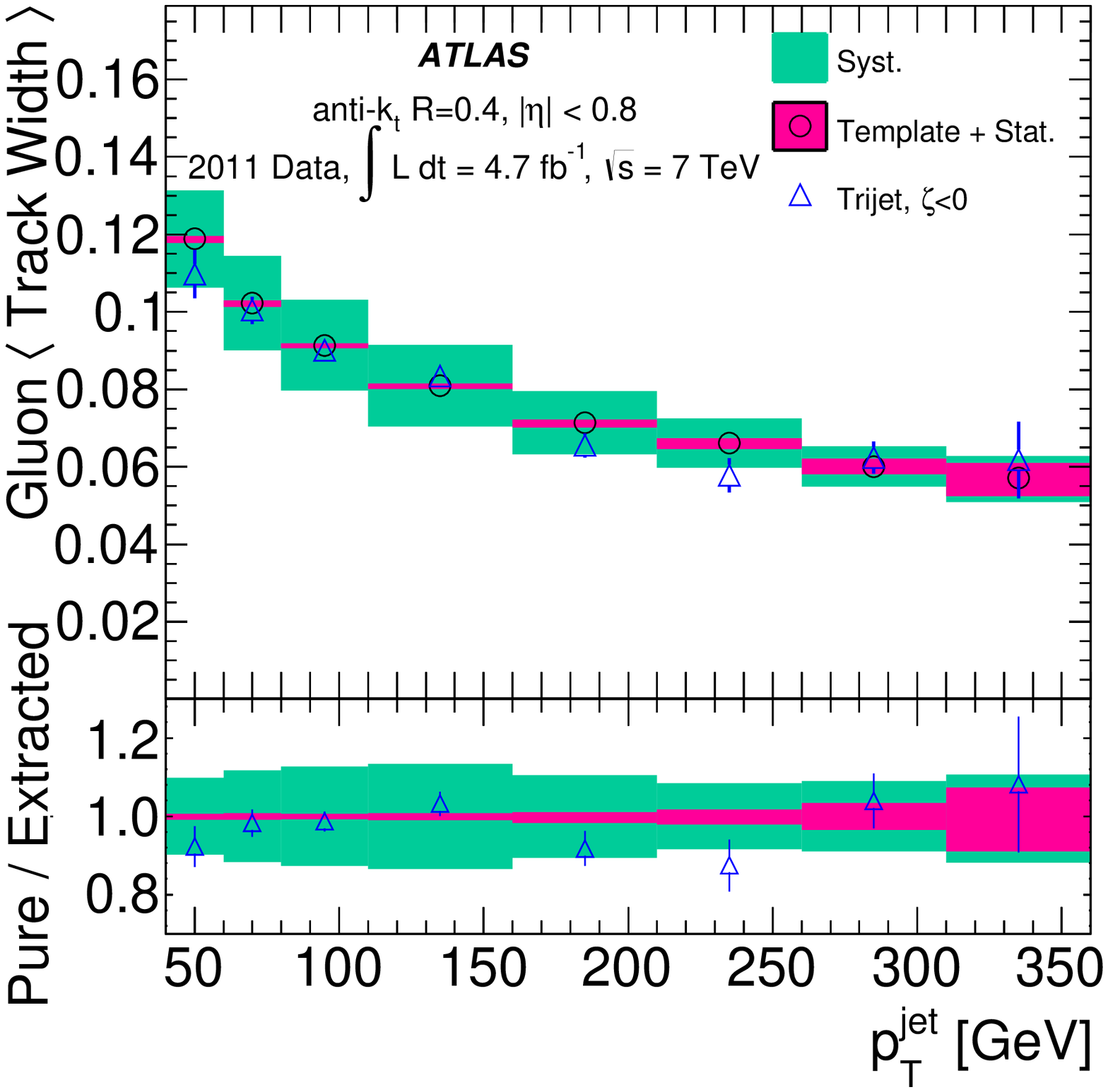}
} \\
\subfigure[]{
\label{fig:pure_trijet_ntrkTrkWidth3}
\includegraphics[width=0.45\textwidth]{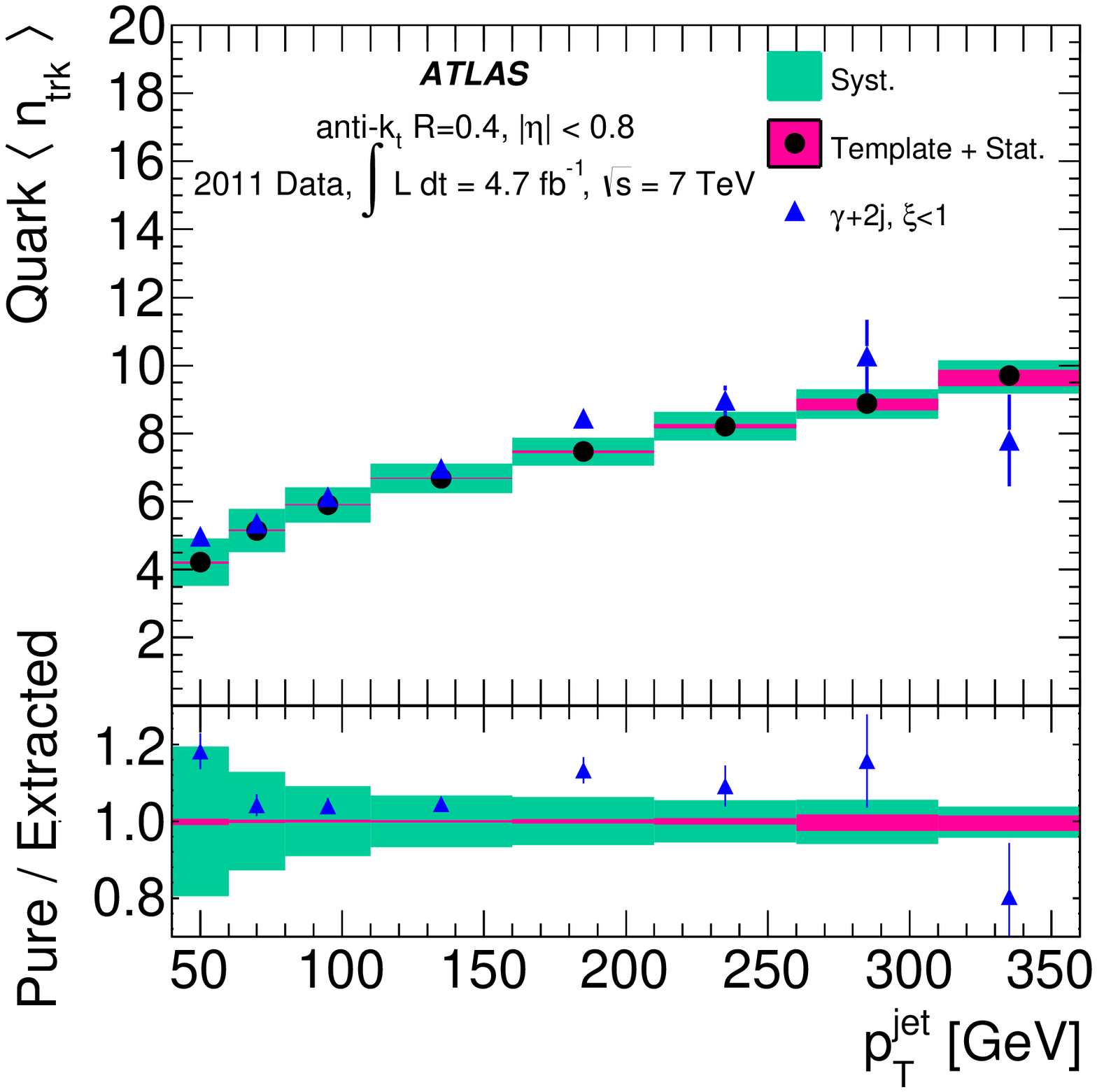}
}
\subfigure[]{
\label{fig:pure_trijet_ntrkTrkWidth4}
\includegraphics[width=0.45\textwidth]{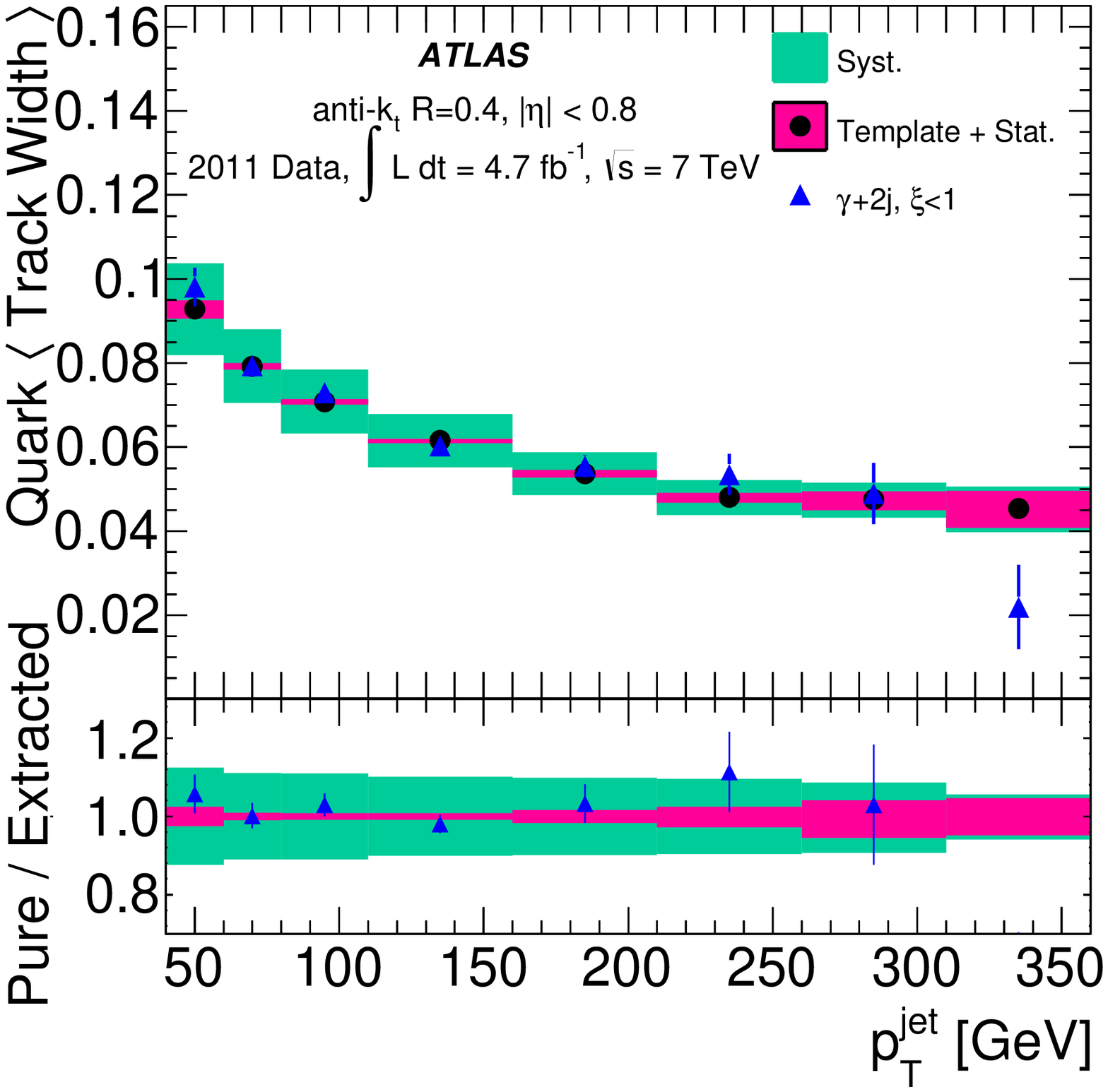}
}
\caption{Top, the jet (a) $n_\text{trk}$ and (b) track width as a function of $\pt$ for jets in a
gluon-jet-enriched trijet sample (triangles) compared to gluon-jet extracted templates (circles) for
$|\eta|<0.8$. Bottom, the jet (c) $n_\text{trk}$ and (d) track width as a function of $\pt$ for jets in a quark-jet-enriched $\gamma$+jet
sample (triangles) compared to quark-jet extracted templates (circles) for jets with $|\eta|<0.8$.
Jets are reconstructed with the anti-$k_t$ algorithm with $R=0.4$. The bottom
panels of the figures show the ratios of the results found in the enriched sample to the extracted
results. Error bars on
the points for the enriched sample correspond to statistical uncertainties. The inner
shaded band around the circles and in the ratio represents statistical uncertainties on the
extracted results, while the outer error band represents the combined systematic and statistical
uncertainties.
}
\label{fig:pure_trijet_ntrkTrkWidth}
\end{center}
\end{figure*}

To evaluate the modelling of gluon-jet properties, events in data with $\zeta<0$ are
compared to those extracted using the template technique described in Sec.~\ref{sec:mcTemp}.
The track multiplicity and jet width are shown in Figs.~\ref{fig:pure_trijet_ntrkTrkWidth1}
and~\ref{fig:pure_trijet_ntrkTrkWidth2}. The
mean values of properties obtained using the purified and (regular) mixed samples generally agree within
statistical and systematic uncertainties. Systematic uncertainties in this figure are calculated as
detailed in Sec.~\ref{sec:systematicsExtr}, and symmetrised around the central value.

\subsection{Validation of quark-jet properties}\label{sec:quarkEnrichmentGamma}

Events containing photons are widely used as an enriched sample of quark-jets.  By selecting
events with photons produced in association with exactly one jet, a sample of quark-jets
that is up to 80\% pure for jets with $\pt>150$~\GeV\ can be constructed.  Although the further
enrichment of quark-jets in this sample is difficult, it is possible to obtain higher purities using
events with a photon and two jets~\cite{schwartz2}. If no other selection cuts are applied, these events have a lower
quark-jet fraction than inclusive $\gamma$-jet production. However, a kinematic selection
can help to identify jets seeded by the parton that is most likely to have radiated the photon.
As that parton must have had electric charge, selecting these jets enhances the purity of
quark-jets and rejects gluon-jets.

Following Ref.~\cite{schwartz2}, a variable is defined that allows the kinematic separation of quark-jets and gluon-jets:

\begin{equation*}
\xi=\eta_\text{jet~1}\times\eta_\gamma+\Delta R_{(\text{jet~2},\gamma)},
\end{equation*}

\noindent where $\eta_\gamma$ ($\eta_\text{jet~1}$) is the $\eta$ of the photon (leading jet), and
$\Delta R_{(\text{jet~2},\gamma)}$ gives the difference in $\eta$--$\phi$ space between the sub-leading
jet and the photon. By imposing a requirement on this variable, purities over 90\% can be achieved,
although with a significant loss of events.

To evaluate the modelling of quark-jet properties, events with $\xi<1$ are compared in data
with those extracted using the template technique described in Sec.~\ref{sec:mcTemp}.
The track multiplicity and jet width are shown in Figs.~\ref{fig:pure_trijet_ntrkTrkWidth3}
and~\ref{fig:pure_trijet_ntrkTrkWidth4}.
The two sets of data agree within statistical and systematic uncertainties.
These results also hold in higher $|\eta|$ bins and for jets reconstructed with the anti-$k_t$
algorithm with $R=0.6$.

Additionally, the production of a \Wboson\ boson in association with a jet can be used to provide a
relatively pure sample of quark-jets. A useful variable in constructing the sample is the jet ``charge'', defined as

\begin{equation*}
c_j = \frac{ \sum_i q_i \times \left| \vec{p_i} \cdot \hat{j} \right|^{1/2} } { \sum_i \left| \vec{p_i} \cdot \hat{j} \right|^{1/2} } 
\end{equation*}

\noindent where the sums run over all tracks associated with the jet, $\hat{j}$ is a unit three-vector
pointing in the direction of the jet momentum, $\vec{p_i}$ is the track momentum three-vector, and
$q_i$ is the track charge. This variable has been found to be useful in discriminating jets originating
from positively charged quarks from those originating from negatively charged
quarks~\cite{JetCharge1,JetCharge2,ATLASttbar}. The leading contribution to \Wboson\ production
results in a jet with charge opposite to that of the \Wboson\ boson. The main backgrounds are from
gluon-jets, including those in events with jets misidentified as leptons, which should have a
charge distribution
that is approximately Gaussian and centred at zero~\footnote{This is not quite the case, as the
initial state at the LHC is more often positively charged than negatively charged.}.

A pure sample of \Wboson\ events, selected as described in Sec.~\ref{sec:data-selection}, is divided
into events in which the leading jet has a charge with the same sign as the identified lepton (SS) and
those in which the charge is opposite (OS). Templates are then constructed for jet properties in the SS
and OS samples, and the SS sample is used to subtract the gluon-jet contribution from the OS template.

\begin{figure}[htbp]
\begin{center}
\subfigure[]{
\includegraphics[width=0.45\textwidth]{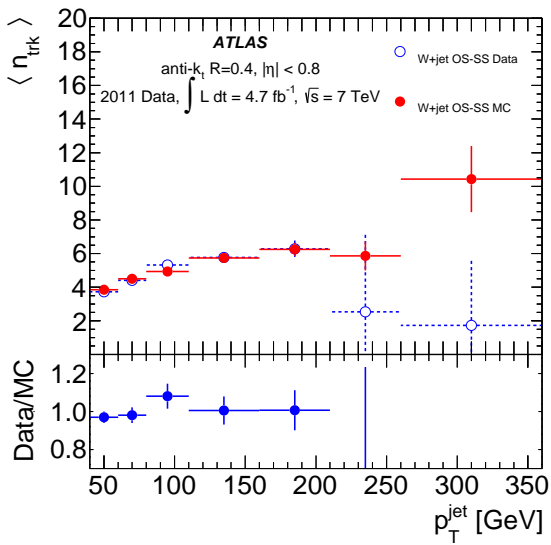}
}
\subfigure[]{
\includegraphics[width=0.45\textwidth]{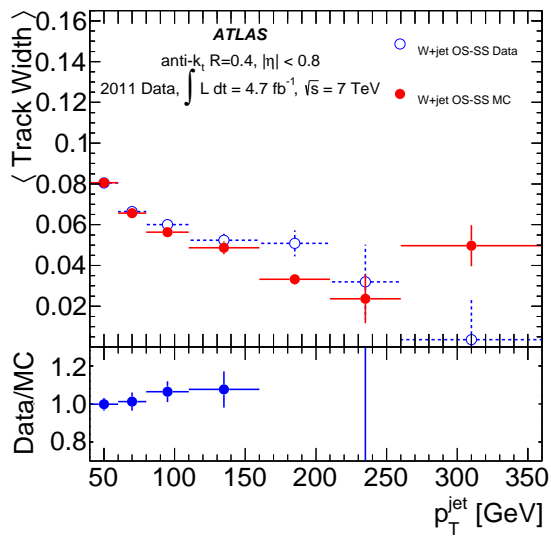}
}
\caption{The jet (a) $n_\text{trk}$ and (b) track width as a function of $\pt$ for quark-jets in
an OS minus SS \Wboson+jet sample (see text) for $|\eta|<0.8$ in {\sc Pythia} 6 MC simulation and in
data. The panels show the ratio of the results in data to those in MC simulation.}
\label{fig:wjetdatamc_ntrk}
\end{center}
\end{figure}

Comparisons between the mean of the OS minus SS distributions in data and MC simulation are
shown in Fig.~\ref{fig:wjetdatamc_ntrk}. The data show reasonable agreement with the
MC simulation, generally within the statistical uncertainties.
The points on these curves disagree at the 10\% level with extracted or purified quark-jet results
shown in previous figures due to a non-closure effect in the method observed at low $\pt$ in the
MC simulation. Results from the $\Wboson+$1-jet MC simulation using generator-based labelling are in agreement
with the quark-jet results from the dijet samples shown in Fig.~\ref{fig:ntrkWidthClos}.

\section{Light-quark/gluon tagger construction and performance}
\label{sec:perfDetails}


The discriminant for quark- and gluon-jets is based on a simple likelihood ratio
that uses the two-dimensional extracted distributions of $n_{\rm trk}$ and track width for quark-
and gluon-jets:

\begin{equation*}
L=\frac{q}{q+g},
\end{equation*}

\noindent where $q$ ($g$) represents the normalised two-dimensional distribution for quark-jets
(gluon-jets). A selection on $L$ is used
in each bin to discriminate quark- and gluon-jets. This discriminant
is built in bins of jet $\pt$ and $\eta$. The two-dimensional distributions are first smoothed using
a Gaussian kernel and then appropriately rebinned to build the discriminant
distribution in such a way that all bins are populated sufficiently.

The performance of the tagger is determined using the two-dimensional extracted distributions
of $n_{\rm trk}$ and track width in data and those obtained for labelled jets in MC simulations.
Systematic uncertainties on the evaluated performance are estimated using alternative templates
as described in Sec.~\ref{sec:systematicsExtr}.
Table~\ref{tab:perfSummary} summarises this performance for jets with $|\eta|<0.8$.
The efficiencies for gluon-jets and quark-jets are evaluated
only at certain operating points with fixed light-quark efficiency. Statistical uncertainties are
evaluated using pseudoexperiments. Systematic
uncertainties are combined in quadrature and affect both the quark- and
gluon-jet efficiency in data. Large differences between MC simulation and data in
the variables used translate into large scale factors in the gluon-jet efficiency. Practically, 
analyses using this tagger would apply the appropriate MC tagger to MC simulation and the data
tagger to data. These scale factors are needed for each MC tagger to create event weights for 
the MC simulation, so that the efficiency in the MC simulation matches the measured efficiency 
in such analyses. Three representative $\pt$ bins are shown in the table.

\begin{table*}[htbp]
\caption{
Summary of the performance of the quark-jet tagger on quark- and gluon-jets in data and {\sc Pythia}6 MC simulation for jets built with the anti-$k_t$ algorithm with $R=0.4$ and with $|\eta|<0.8$.
The first error corresponds to the statistical uncertainty, while the second corresponds to the systematic uncertainty.
The scale factor is the ratio of data to MC simulation.
\label{tab:perfSummary}}
\begin{center}
\begin{tabular}{|c|l|l|l|l|l|l|}
\hline
& \multicolumn{2}{|c|}{Monte Carlo} & \multicolumn{2}{|c|}{Data} & \multicolumn{2}{|c|}{Scale Factor} \\
\hline
& $\epsilon_{\rm quark}$ & $\epsilon_{\rm gluon}$ & $\epsilon_{\rm quark}$ & $\epsilon_{\rm gluon}$ & SF$_{\rm quark}$ & SF$_{\rm gluon}$ \\
\hline\noalign{\smallskip}
\multirow{4}{*}{\begin{sideways}$\pt=60$--$80 \GeV$\end{sideways}}
& 30\% &  8.4\% &  ($30.0\pm0.8^{+3.2}_{-5.3}$)\%    &  ($11.9\pm0.3^{+7.5}_{-2.9}$)\%    &  $1.00\pm0.03^{+0.11}_{-0.18}$ & $1.42\pm0.04^{+0.89}_{-0.34}$ \\&&&&&&\\
& 50\% & 21.0\% &  ($50.0^{+1.4+4.3}_{-1.3-6.8}$)\%  &  ($26.6^{+0.8+7.1}_{-0.6-3.9}$)\%  &  $1.00^{+0.027+0.09}_{-0.026-0.14}$ & $1.27^{+0.04+0.34}_{-0.03-0.19}$ \\&&&&&&\\
& 70\% & 41.5\% &  ($70.0^{+1.7+3.9}_{-1.5-11.0}$)\%  &  ($48.4^{+1.1+4.7}_{-0.9-6.0}$)\%  &  $1.00^{+0.024+0.06}_{-0.022-0.16}$ & $1.17^{+0.03+0.11}_{-0.02-0.14}$ \\&&&&&&\\
& 90\% & 69.9\% &  ($90.0^{+1.5+1.7}_{-1.3-3.3}$)\%  &  ($80.2^{+1.0+5.6}_{-0.8-2.2}$)\%   &  $1.00^{+0.02+0.02}_{-0.01-0.04}$ & $1.15^{+0.015+0.08}_{-0.012-0.03}$ \\&&&&&&\\
\hline\noalign{\smallskip}
\multirow{4}{*}{\begin{sideways}\centering$\pt=110$--$160 \GeV$\end{sideways}}
& 30\% &  5.7\% &  ($30.0\pm0.6^{+2.8}_{-4.6}$)\%  & ($11.6^{+0.6+6.2}_{-0.4-4.6}$)\%   & $1.00\pm0.02^{+0.09}_{-0.15}$ & $2.03^{+0.11+1.08}_{-0.08-0.81}$ \\ &&&&&&\\
& 50\% & 13.9\% &  ($50.0\pm1.0^{+4.1}_{-6.1}$)\%  & ($24.3^{+1.2+7.4}_{-0.8-9.2}$)\%   & $1.00\pm0.02^{+0.08}_{-0.12}$ & $1.75^{+0.09+0.53}_{-0.06-0.66}$ \\ &&&&&&\\
& 70\% & 29.7\% &  ($70.0^{+1.0+3.9}_{-1.1-8.5}$)\%  & ($45.3^{+1.5+4.6}_{-1.1-9.3}$)\% & $1.00^{+0.01+0.06}_{-0.02-0.12}$ & $1.52^{+0.05+0.15}_{-0.04-0.31}$ \\ &&&&&&\\
& 90\% & 64.8\% &  ($90.0^{+0.5+2.0}_{-0.6-2.6}$)\%  & ($78.1^{+1.0+3.5}_{-0.6-6.0}$)\% & $1.00^{+0.006+0.02}_{-0.007-0.03}$ & $1.21^{+0.02+0.05}_{-0.01-0.09}$ \\ &&&&&&\\
\hline\noalign{\smallskip}
\multirow{4}{*}{\begin{sideways}\centering$\pt=310$--$360 \GeV$\end{sideways}}
& 30\% &  3.9\% & ($30.0^{+5.0+2.1}_{-7.1-4.7}$)\% & ($11^{+5+8}_{-7-4}$)\% & $1.00^{+0.17+0.07}_{-0.24-0.16}$ & $2.8^{+1.4+2.0}_{-1.9-1.1}$ \\&&&&&&\\
& 50\% & 10.3\% & ($50.0^{+8.1+3.0}_{-11.6-8.3}$)\%& ($23^{+10+8}_{-12-9}$)\%& $1.00^{+0.16+0.06}_{-0.23-0.17}$ & $2.2^{+1.0+0.8}_{-1.1-0.9}$ \\&&&&&&\\
& 70\% & 23.5\% & ($70.0^{+7.2+3.1}_{-8.8-7.0}$)\% & ($43^{+8+6}_{-12-10}$)\% & $1.00^{+0.10+0.04}_{-0.13-0.10}$ & $1.81^{+0.35+0.23}_{-0.51-0.42}$ \\&&&&&&\\
& 90\% & 58.9\% & ($90.0^{+5.0+1.8}_{-4.9-3.1}$)\% & ($80^{+6+4}_{-10-7}$)\% & $1.00^{+0.06+0.02}_{-0.05-0.03}$ & $1.37^{+0.10+0.07}_{-0.17-0.11}$ \\&&&&&&\\
\hline
\end{tabular}
\end{center}
\end{table*}

The difference in efficiency between data and MC simulation is particularly large for
the tightest operating point at high $\pt$.  It improves for the loosest operating points
and is generally better for the lowest $\pt$ bins.
The efficiencies extracted from data show a much weaker dependence on $\pt$ than is suggested
by {\sc Pythia} 6. No strong dependence on $\eta$ is observed in any sample. The
performance obtained here in {\sc Pythia} 6 compares well with the generator-level studies
presented in Ref.~\cite{schwartz1}. The systematic uncertainties are dominated by the
uncertainty due to the sample dependence.

\begin{figure}[!htp]
\begin{center}
\subfigure[]{
\includegraphics[width=0.45\textwidth]{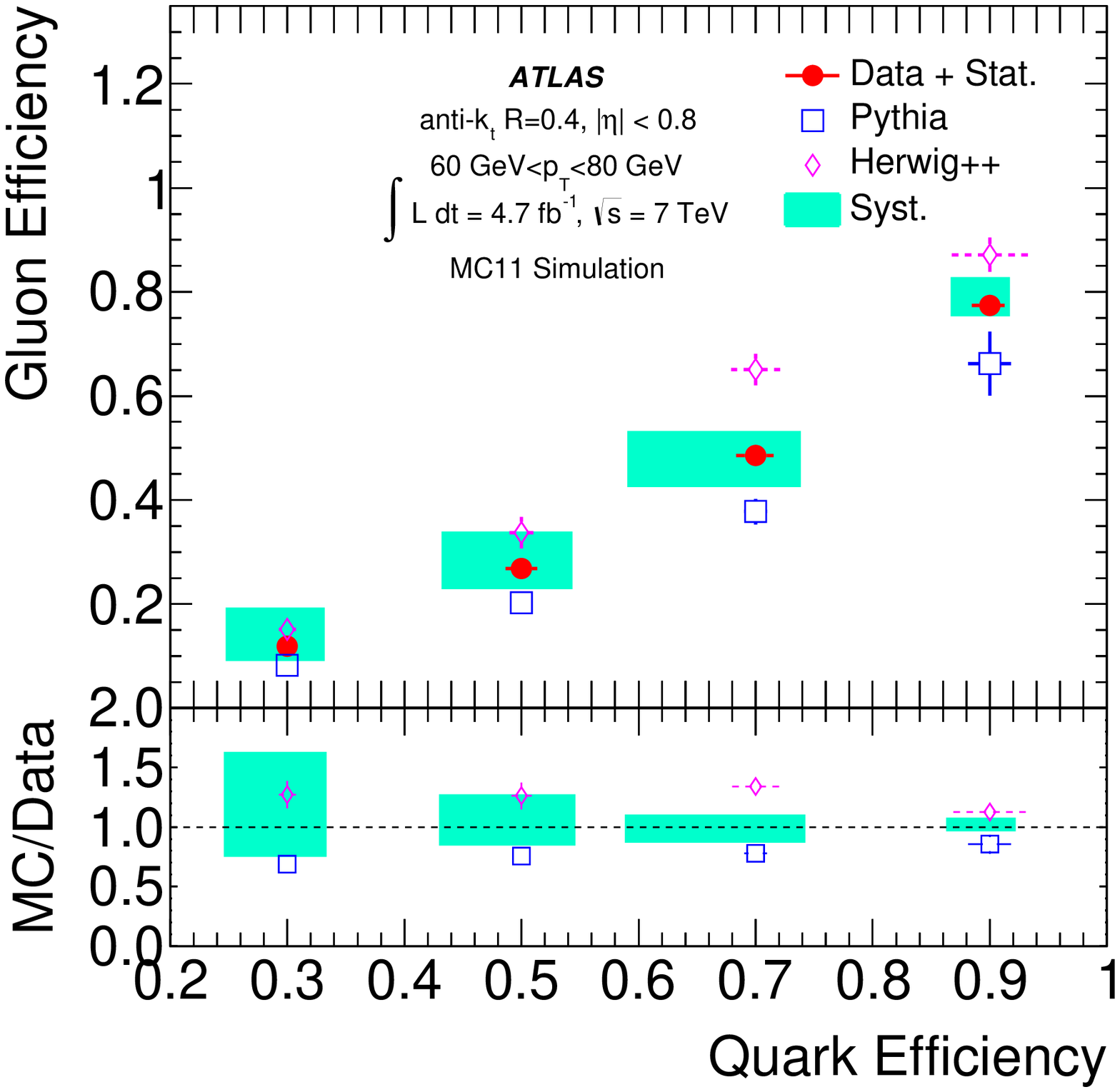}
\label{fig:perf_summary_a}
}
\subfigure[]{
\includegraphics[width=0.45\textwidth]{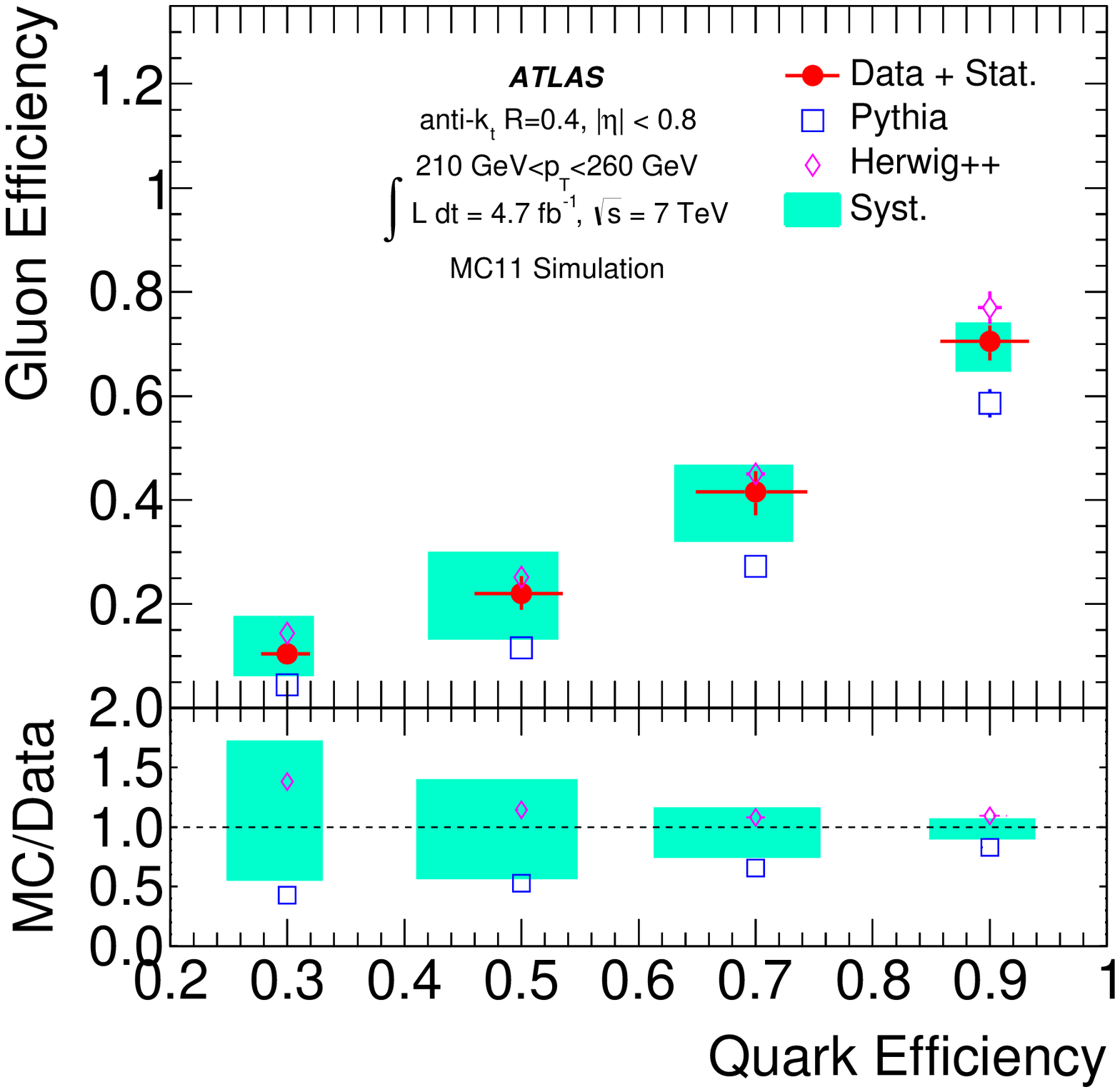}
\label{fig:perf_summary_b}
}
\caption{ Gluon-jet efficiency as a function of quark-jet efficiency calculated using jet
properties extracted from data (solid symbols) and from MC-labelled jets
from the dijet {\sc Pythia} 6 (empty squares) and {\sc Herwig}++ (empty diamonds) samples.
Jets with (a) $60<\pt<80$~\GeV\ and (b) $210<\pt<260$~\GeV\  and $|\eta|<0.8$
are reconstructed with the anti-$k_t$ algorithm with $R=0.4$.
The shaded band shows the total systematic uncertainty on the data. The bottom of the plot
shows the ratios of each MC simulation to the data. The error bands on the performance
in the data are drawn around 1.0.
}
\label{fig:perf_summary}
\end{center}
\end{figure}

The efficiencies of the tagger in MC simulation and in data are summarised in Fig.~\ref{fig:perf_summary},
where the performance estimated from labelled jets in dijet MC simulations and extracted data are shown.
Two MC simulation-based taggers were used to produce this figure, one developed using distributions
extracted in {\sc Pythia} 6, which is applied to the {\sc Pythia} 6 samples, and another derived
from {\sc Herwig}++, used for the {\sc Herwig}++ samples.
As expected from Sec.~\ref{sec:extraction}, the data do not agree well with either
{\sc Pythia} 6 or  {\sc Herwig}++. Differences between data and {\sc Pythia} 6 are within
systematic uncertainties at low $\pt$, but are more significant at high $\pt$ for those points
for which a large sample is available in the data. The tagger performs worse in {\sc Herwig}++
than on data at low $\pt$ (\ref{fig:perf_summary_a}), but there is fair agreement in its performance
for high $\pt$ jets (\ref{fig:perf_summary_b}).
Comparable results are observed for higher $|\eta|$ ranges, but with larger
statistical uncertainties.

\begin{figure}[!htp]
\begin{center}
\subfigure[]{
\includegraphics[width=0.45\textwidth]{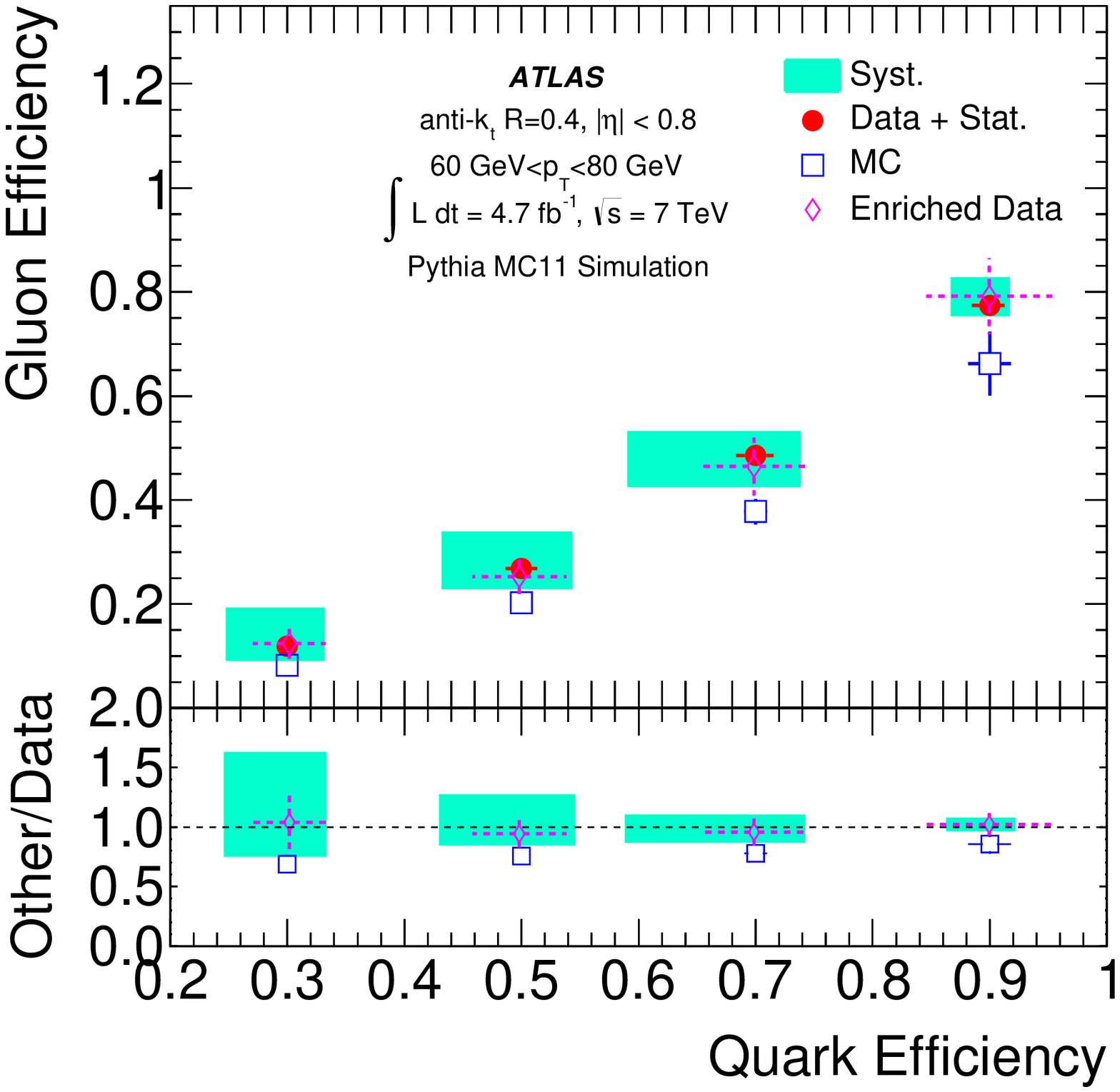}
\label{fig:perf_summary_pur_a}
}
\subfigure[]{
\includegraphics[width=0.45\textwidth]{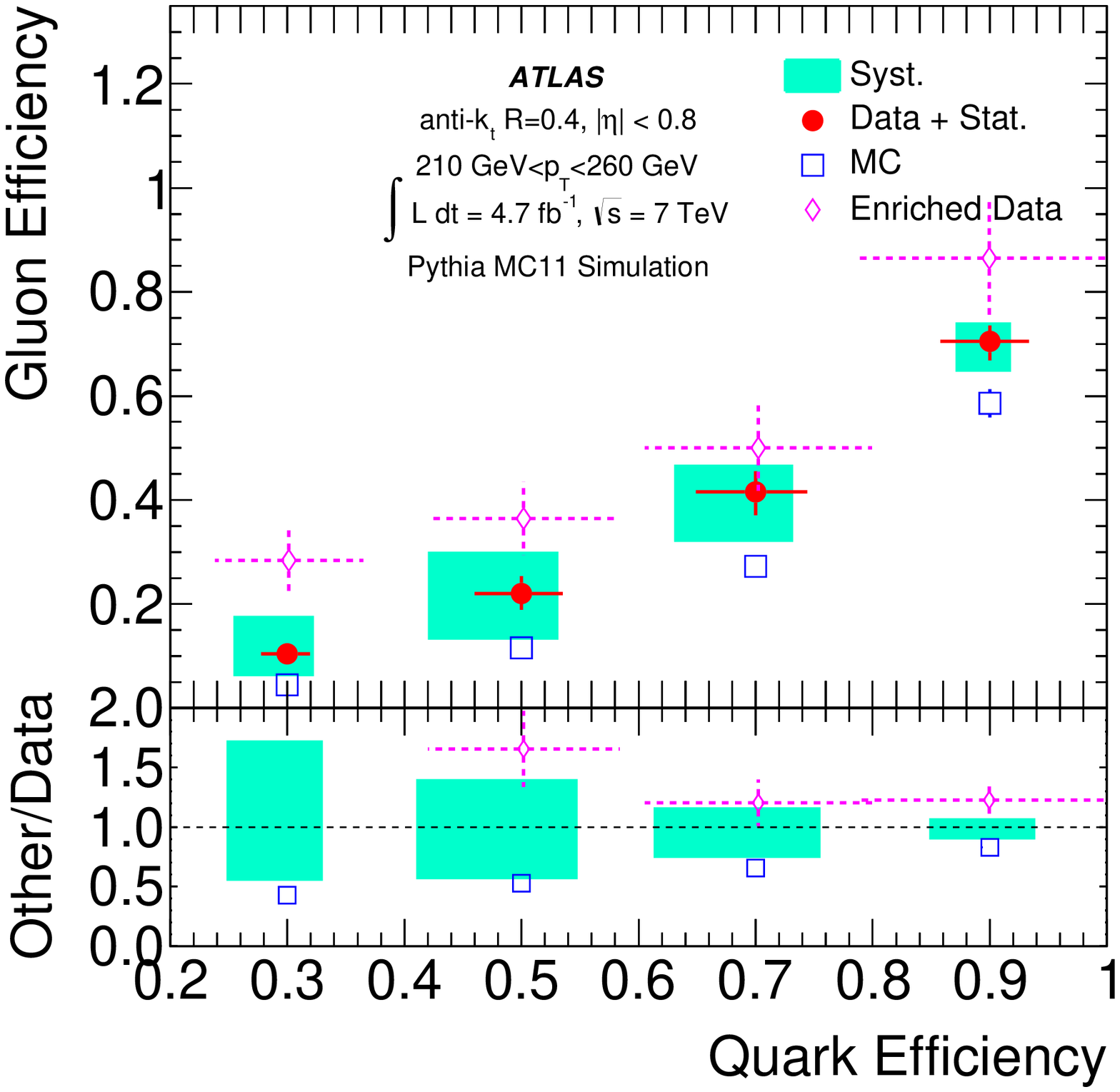}
\label{fig:perf_summary_pur_b}
}
\caption{ Gluon-jet efficiency as a function of quark-jet efficiency as calculated using jet properties extracted
from data (solid symbols), purified in data through kinematic cuts (empty diamonds), and extracted from
{\sc Pythia} 6 MC simulation (empty squares).
Jets with (a) $60<\pt<80$~\GeV\ and (b) $210<\pt<260$~\GeV\ and $|\eta|<0.8$ are reconstructed with the anti-$k_t$ algorithm with $R=0.4$.
The shaded band shows the total systematic uncertainty on the data.
The bottom of the plot
shows the ratio of {\sc Pythia} 6 MC simulation or the enriched data samples
to the extracted data. The error bands on the performance
in the data are drawn around 1.0.
}
\label{fig:perf_summary_pur}
\end{center}
\end{figure}

\begin{figure}[!htp]
\begin{center}
\subfigure[]{
\includegraphics[width=0.45\textwidth]{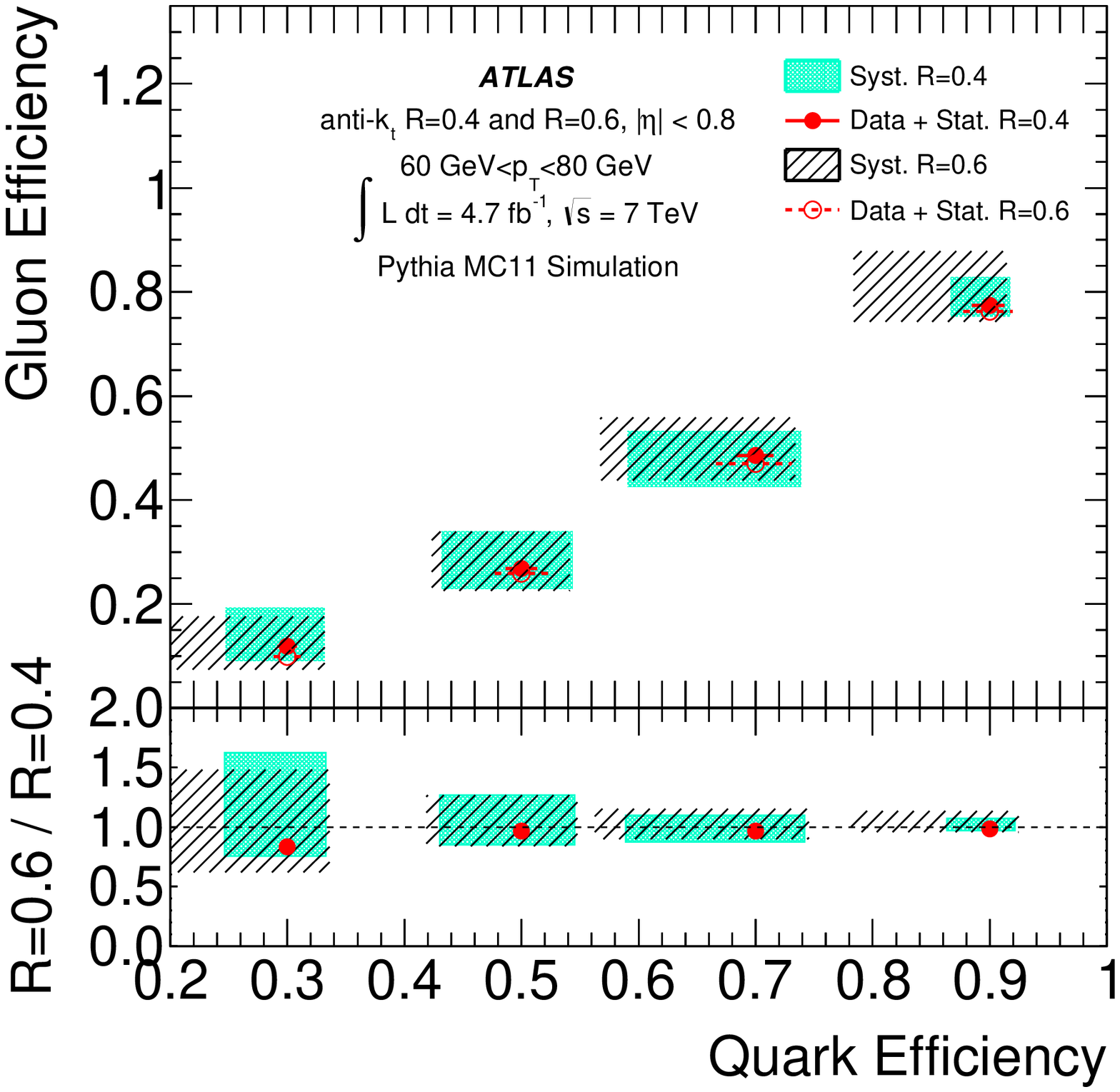}
}
\subfigure[]{
\includegraphics[width=0.45\textwidth]{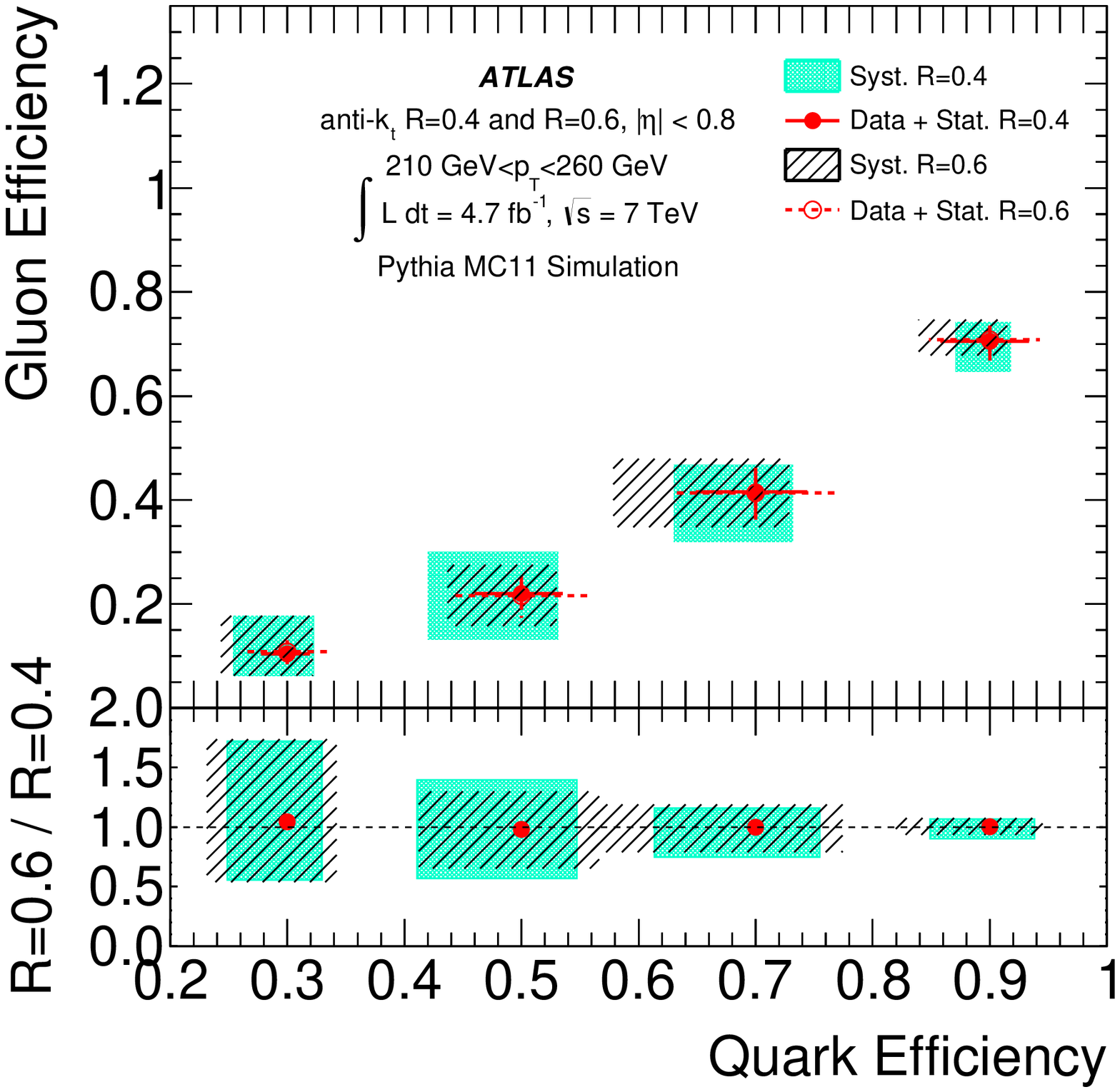}
}
\caption{ Gluon-jet efficiency as a function of quark-jet efficiency as calculated using extracted
jet properties for jets with (a) $60<\pt<80$~\GeV\ and (b) $210<\pt<260$~\GeV\ and $|\eta|<0.8$ reconstructed with
the anti-$k_t$ algorithm with $R=0.4$ (solid symbols) and $R=0.6$ (empty symbols).
The shaded (hashed) band represents the total systematic uncertainty on the $R=0.4$ ($R=0.6$) data points. When
hardly visible, the empty symbols are just behind the solid symbols.
The bottom of the plot
shows the ratio of the performance in data obtained for $R=0.6$ to that for $R=0.4$. Error bands
are drawn around 1.0.
}
\label{fig:perf_summary_rsize}
\end{center}
\end{figure}

The performance can also be calculated using the relatively pure samples obtained in
trijet and $\gamma+$2-jet events (see Sec.~\ref{sec:disc}).
The efficiencies obtained using purified samples are compared in Fig.~\ref{fig:perf_summary_pur} to those obtained
using the extracted discriminant distribution. The agreement within
systematic uncertainties, particularly in Fig.~\ref{fig:perf_summary_pur_a}, further validates the extraction method.
Some small differences, like those in Fig.~\ref{fig:perf_summary_pur_b}, should be expected from impurities in the
quark and gluon purified samples.
A comparison of performance in jets with radius parameters of $R=0.4$ and $R=0.6$ in data
and MC simulation is shown in Fig.~\ref{fig:perf_summary_rsize}.  The performance is
comparable with the two jet sizes.

\section{Conclusions}
\label{sec:conclusions}


Several variables that are sensitive to differences between quark- and gluon-jets
were studied in various MC simulations and 4.7~fb$^{-1}$ of $\sqrt{s}=7$~\TeV\ 
$pp$ collision data collected with the ATLAS detector at the LHC during the year 2011.
Two of these variables, chosen to be relatively weakly correlated and stable against pile-up,
were used to build a likelihood-based discriminant to select quark-jets and
reject gluon-jets. Because of non-negligible differences in these variables between
data and MC simulations, a data-driven technique was developed to extract
the discriminant from the data and the MC simulations independently. This technique
exploits significant, $\pt$ dependent differences in the quark- and gluon-jet content
between dijet and $\gamma+$jet samples.

A detailed study of the jet properties reveals that
quark- and gluon-jets look more similar to each other in the data than in the {\sc Pythia}~6
simulation and less similar than in the {\sc Herwig}++ simulation. As a result,
the ability of the tagger to reject gluons at a fixed quark efficiency is up to a factor
of two better in {\sc Pythia} 6 and up to 50\% worse in {\sc Herwig}++ than in data.
Gluon-jet efficiencies in data of $\approx 11\%$ and 80\% are achieved for quark-jet efficiencies of
$\approx 30\%$ and 90\%, respectively. Relative uncertainties of $\approx 5-50\%$ ($\approx 3-20\%$)
were evaluated for the estimate of these gluon-jet (quark-jet) efficiencies,
with the uncertainties increasing for operating points with lower quark-jet efficiency.
These uncertainties are dominated by differences in the properties of quark- and gluon-jets
in the calibration samples (dijet and $\gamma+$jet) and are potentially caused by
effects such as colour flow, which can make radiation around jets different for jets
in different samples, even if they share the same partonic origin. These differences are predicted
to be of different magnitude by the two generators studied and, without further insight, prevent
final-state-dependent taggers to be developed. The differences between the properties in the two
samples are typical of the variations of the properties observed in other samples studied.

The likelihood-based discriminants were
studied independently in kinematically purified gluon-jet and quark-jet samples in data.
Agreement is found within systematic uncertainty between the properties that are used to build the
discriminant for the pure samples and the mixed samples.  The same is true for the tagger efficiencies.

Because their properties differ, the same likelihood-ratio discriminant cannot be applied to non-isolated
jets.  However, using the methodology described in this paper, a discriminant for non-isolated
jets with typical rejections and efficiencies comparable to those of the isolated-jet
discriminant can be derived.

\appendix

\section{Performance of EEC angularities}
\label{sec:EECresults}

\begin{figure}[htp]
\begin{center}
\subfigure[]{
\includegraphics[width=0.47\textwidth]{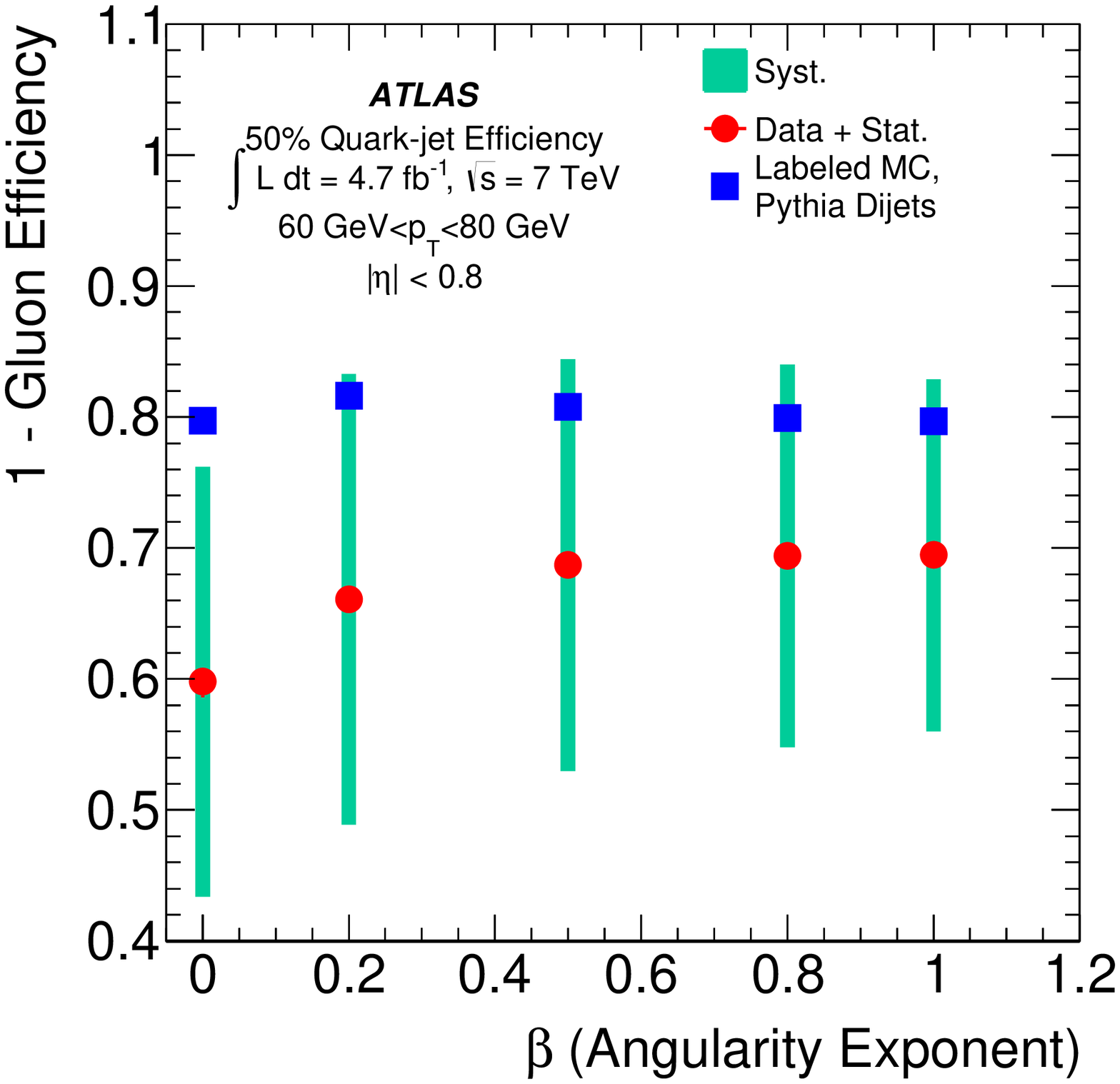}
}
\subfigure[]{
\includegraphics[width=0.47\textwidth]{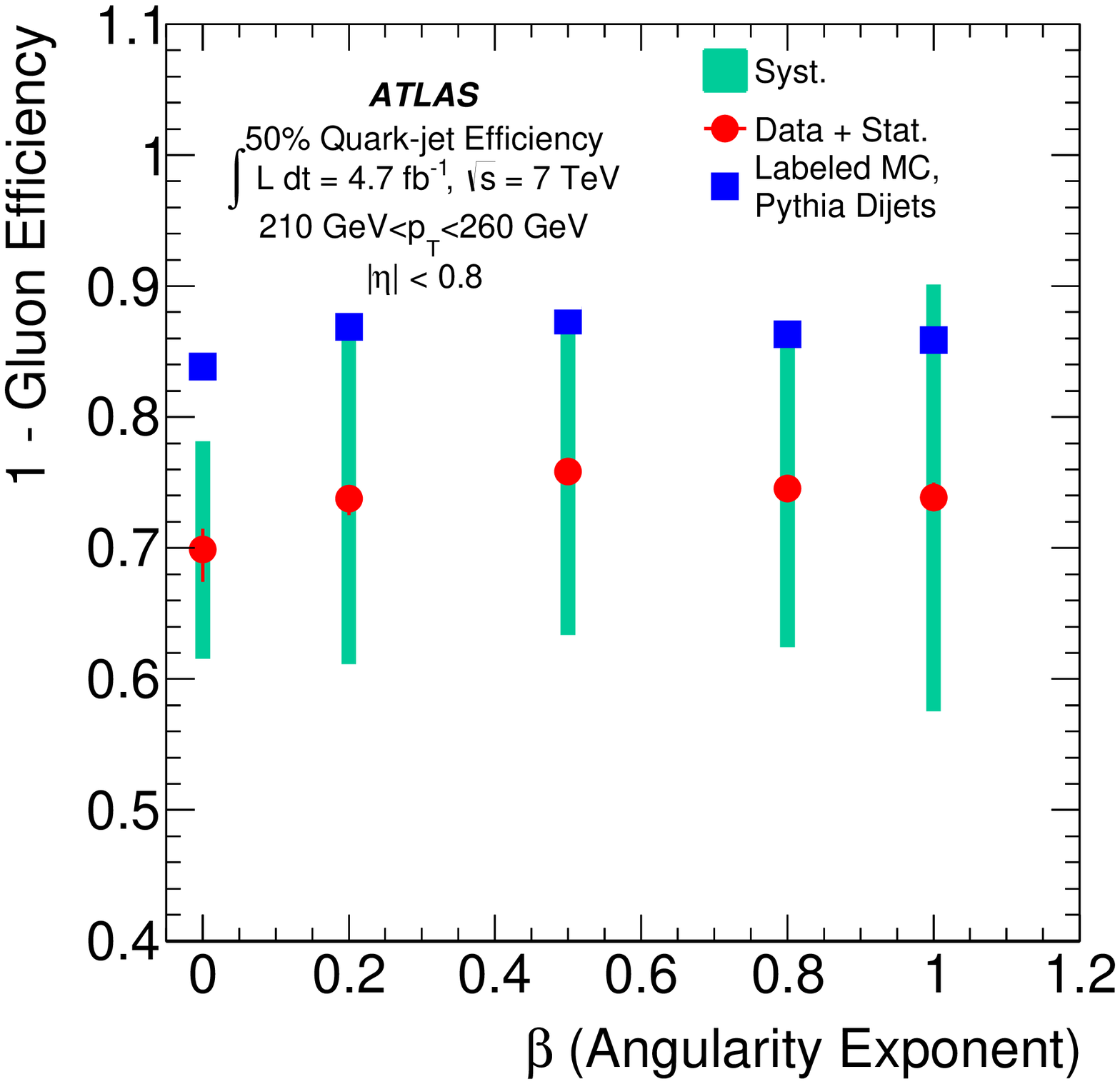}
}
\caption{
One minus the gluon-jet efficiency (for comparison with Ref.~\cite{Larkoski2013eya}) as a function of
$\beta$ (see Sec.~\ref{sec:variables}) for EEC angularities, calculated using extracted jet properties from
data (solid circles) and dijet {\sc Pythia} 6 MC simulation (solid squares) for jets with (a) $60<\pt<80$~\GeV\
and (b) $210<\pt<260$~\GeV\ and $|\eta|<0.8$ reconstructed with the anti-$k_t$ algorithm with $R=0.4$.
The shaded bands represent the total systematic uncertainty on the data points.
}
\label{fig:angSummary}
\end{center}
\end{figure}

The EEC angularities described in Sec.~\ref{sec:variables} include a free parameter $\beta$
that affects the performance of the variables for quark/gluon discrimination.  Recent
studies~\cite{Larkoski2013eya} suggest that smaller values of the exponent $\beta$
provide stronger gluon-jet rejection for the same quark-jet efficiency.  Figure~\ref{fig:angSummary}
shows one minus the gluon-jet efficiency (for comparison with Ref.~\cite{Larkoski2013eya}) as a
function of $\beta$, for a fixed quark-jet efficiency of 50\%, in data and {\sc Pythia} 6 MC
simulation.  The MC simulation shows a weak dependence on $\beta$, with
optimal performance for a $\beta$ value between 0.2 and 0.4. A similar trend is observable
in data at high jet $\pt$.
At low jet
$\pt$, however, the performance falls off with lower $\beta$, with the highest few $\beta$ points
showing comparable performance.  The worst performance at all $\pt$ is given by $\beta=0$, which
uses the $\pt$ of the tracks without angular information, emphasising the high-$\pt$ tracks at the
core of the jet for which the tracking momentum resolution is worse and inefficiencies and fakes due to shared
hits between tracks in the detector become more common. At high jet $\pt$, the point at $\beta=0$ shows reduced systematic uncertainties with
respect to many of the other points, though the sample dependence is still significant.

It should be
noted that the dependence on $\beta$ resembles more closely that found in Ref.~\cite{Larkoski2013eya}
when using quark- and gluon-jets from either the dijet or $\gamma+$jet samples exclusively in building
and testing the tagger in MC simulation. The template method developed in this paper is
sensitive to the sample dependence of these variables. Since quark- and gluon-jets from the two samples
used by the method show differences in these variables, the method is
not capable of distinguishing this trend.
The significant uncertainties on the data points in Fig.~\ref{fig:angSummary}
are mostly from the uncertainties associated with this sample dependence.
This serves
to emphasise the importance of data-based validation of quark-jet/gluon-jet discriminants, as MC
simulation may not correctly describe the jet properties observed in data, as well as the importance
of correct MC event generator tunes that describe the jet properties and their potential sample dependence.

\section*{Acknowledgements}

We thank CERN for the very successful operation of the LHC, as well as the
support staff from our institutions without whom ATLAS could not be
operated efficiently.

We acknowledge the support of ANPCyT, Argentina; YerPhI, Armenia; ARC,
Australia; BMWF and FWF, Austria; ANAS, Azerbaijan; SSTC, Belarus; CNPq and FAPESP,
Brazil; NSERC, NRC and CFI, Canada; CERN; CONICYT, Chile; CAS, MOST and NSFC,
China; COLCIENCIAS, Colombia; MSMT CR, MPO CR and VSC CR, Czech Republic;
DNRF, DNSRC and Lundbeck Foundation, Denmark; EPLANET, ERC and NSRF, European Union;
IN2P3-CNRS, CEA-DSM/IRFU, France; GNSF, Georgia; BMBF, DFG, HGF, MPG and AvH
Foundation, Germany; GSRT and NSRF, Greece; ISF, MINERVA, GIF, I-CORE and Benoziyo Center,
Israel; INFN, Italy; MEXT and JSPS, Japan; CNRST, Morocco; FOM and NWO,
Netherlands; BRF and RCN, Norway; MNiSW and NCN, Poland; GRICES and FCT, Portugal; MNE/IFA, Romania; MES of Russia and ROSATOM, Russian Federation; JINR; MSTD,
Serbia; MSSR, Slovakia; ARRS and MIZ\v{S}, Slovenia; DST/NRF, South Africa;
MINECO, Spain; SRC and Wallenberg Foundation, Sweden; SER, SNSF and Cantons of
Bern and Geneva, Switzerland; NSC, Taiwan; TAEK, Turkey; STFC, the Royal
Society and Leverhulme Trust, United Kingdom; DOE and NSF, United States of
America.

The crucial computing support from all WLCG partners is acknowledged
gratefully, in particular from CERN and the ATLAS Tier-1 facilities at
TRIUMF (Canada), NDGF (Denmark, Norway, Sweden), CC-IN2P3 (France),
KIT/GridKA (Germany), INFN-CNAF (Italy), NL-T1 (Netherlands), PIC (Spain),
ASGC (Taiwan), RAL (UK) and BNL (USA) and in the Tier-2 facilities
worldwide.

\clearpage

%

\input{atlas_authlist.tex}

\end{document}

%% file: atlas_authlist.tex
\clearpage
\onecolumn

\begin{flushleft}
{\Large The ATLAS Collaboration}

\bigskip

G.~Aad$^{\rm 84}$,
B.~Abbott$^{\rm 112}$,
J.~Abdallah$^{\rm 152}$,
S.~Abdel~Khalek$^{\rm 116}$,
O.~Abdinov$^{\rm 11}$,
R.~Aben$^{\rm 106}$,
B.~Abi$^{\rm 113}$,
M.~Abolins$^{\rm 89}$,
O.S.~AbouZeid$^{\rm 159}$,
H.~Abramowicz$^{\rm 154}$
% ...
\bigskip
\\
% ...
$^{11}$ Institute of Physics, Azerbaijan Academy of Sciences, Baku, Azerbaijan\\
% ...
$^{84}$ CPPM, Aix-Marseille Universit{\'e} and CNRS/IN2P3, Marseille, France\\
% ...
$^{89}$ Department of Physics and Astronomy, Michigan State University, East Lansing MI, United States of America\\
% ...
$^{106}$ Nikhef National Institute for Subatomic Physics and University of Amsterdam, Amsterdam, Netherlands\\
% ...
$^{112}$ Homer L. Dodge Department of Physics and Astronomy, University of Oklahoma, Norman OK, United States of America\\
$^{113}$ Department of Physics, Oklahoma State University, Stillwater OK, United States of America\\
% ...
$^{116}$ LAL, Universit{\'e} Paris-Sud and CNRS/IN2P3, Orsay, France\\
% ...
$^{152}$ Institute of Physics, Academia Sinica, Taipei, Taiwan\\
% ...
$^{154}$ Raymond and Beverly Sackler School of Physics and Astronomy, Tel Aviv University, Tel Aviv, Israel\\
% ...
$^{159}$ Department of Physics, University of Toronto, Toronto ON, Canada\\
% ...
\end{flushleft}